\documentclass[showpacs,preprintnumbers,amsmath,amssymb,superscriptaddress,nofootinbib]{revtex4}

\usepackage[english]{babel}
\usepackage{amsmath}
\usepackage{graphicx}
\usepackage{dcolumn}
\usepackage{bm}
\usepackage{amscd}
\usepackage{epstopdf}
\usepackage{hyperref}
\usepackage{textcomp}
\usepackage{natbib}
\usepackage{color}

\newcommand{\be}{\begin{equation}} 
\newcommand{\ee}{\end{equation}}
\newcommand{\ba}{\begin{eqnarray}}
\newcommand{\ea}{\end{eqnarray}}

\usepackage{lipsum}
\newcommand{\der}[2]{\frac{\partial #1}{\partial #2}}

\newcommand{\dersec}[2]{\frac{\partial^{2} #1}{\partial #2^{2}}}

\newcommand{\dermixd}[3]{\frac{\partial^{2} #1}{\partial #2 \partial #3}}

\newcommand{\av}[1]{\langle #1 \rangle}

\newcommand{\tonde}[1]{\left( #1\right)}
\newcommand{\quadre}[1]{\left[ #1\right]}
\newcommand{\set}[1]{\left\{ #1\right\}}

\newcommand{\A}{{A}}
\newcommand{\B}{{B}}
\newcommand{\X}{{j}}

\begin{document}

\title{Complete integrability of information processing by biochemical reactions}

\author{Elena Agliari}
\email[]{agliari@mat.uniroma1.it}
\affiliation{Dipartimento di Matematica, Sapienza Universit\`a di Roma, Italy}
\affiliation{Istituto Nazionale d'Alta Matematica (GNFM-INdAM), Rome (IT)}

\author{Adriano Barra}
\email[]{adriano.barra@roma1.infn.it}
\affiliation{Department of Computer Science, Sapienza Universit\`a di Roma, Italy}
\affiliation{Istituto Nazionale d'Alta Matematica (GNFM-INdAM), Rome (IT)}

\author{Lorenzo Dello Schiavo}
\email[]{delloschiavo@iam.uni-bonn.de}
\affiliation{Institut f\"ur Angewandte Mathematik, Rheinische Friedrich-Wilhelms-Universit\"at Bonn, Germany}

\author{Antonio Moro}
\email[]{antonio.moro@northumbria.ac.uk}
\affiliation{Department of Mathematics, Physics and Electrical Engineering, University of Northumbria Newcastle, United Kingdom}

\date{\today}

\begin{abstract}
Statistical mechanics provides an effective framework to investigate \emph{information processing} in biochemical reactions. Within such framework far-reaching analogies are established among \emph{(anti-) cooperative collective behaviors} in chemical kinetics, \emph{(anti-)ferromagnetic spin models} in statistical mechanics and \emph{operational amplifiers/flip-flops} in cybernetics.
\newline
The underlying modeling -- based on spin systems -- has been proved to be accurate for a wide class of systems matching classical (e.g. Michaelis--Menten, Hill, Adair) scenarios in the infinite-size approximation.
However, the current research in biochemical information processing has been focusing on systems involving a relatively small number of units, where this approximation is no longer valid.
\newline
Here we show that the whole statistical mechanical description of reaction kinetics can be re-formulated via a mechanical analogy -- based on completely integrable hydrodynamic-type systems of PDEs -- which provides explicit finite-size solutions, matching recently investigated phenomena (e.g. noise-induced cooperativity, stochastic bi-stability, quorum sensing).
\newline
The resulting picture, successfully tested against a broad spectrum of data, constitutes a neat rationale for a \emph{numerically effective} and \emph{theoretically consistent} description of collective behaviors in biochemical reactions.
\end{abstract}

\maketitle

\section*{Introduction}

Since the pioneering work by Hopfield \cite{hopfield2} on kinetic proofreading and the early applications of stochastic techniques to reaction kinetics by Chay and Ho \cite{chay} or Wyman and Phillipson \cite{wyman}, the combination of a number of recent significant results, both experimental (see e.g. \cite{Warren,Ricci,Gardner-Nature2000,Samoilov-PNAS2005}) and theoretical (see e.g. \cite{kardar-bistability,McKane,Bialek-PRL2008,pigo}), has boosted the current understanding of biochemical information-processing systems, namely of how the thermodynamics of biochemical reactions {\em spontaneously} encodes information processing.
\newline
These results stem from investigations scattered over different fields of biological research involving, for instance, inter-cellular \cite{kardar-bistability,Kardar-PNAS2013} and intra-cellular signalling \cite{Bialek-PRL2008,Ricci,winfree3}, enzymatic cycles \cite{Samoilov-PNAS2005}, ribo and toggle switches \cite{zhang2,Warren,Warren2,Warren3,Kim-Wang}, ultra-sensitive mechanisms \cite{zhang1,Noise-Suppressor,bradshaw}, DNA-computing \cite{winfree2,winfree3}, transcriptional and regulatory networks \cite{Leibler-Nature2000,transcriptional-scenario,Tian}, and more.
\newline
The theoretical description of such systems is typically based on stochastic approaches, e.g., Fokker--Planck equations suitably adapted to the cases of interest, leading to the chemical extension of the master equation approach (see e.g. \cite{McKane,chemical-master-equation,Tian} and references therein).

\begin{figure}[tbp]
\noindent \begin{centering}
\includegraphics[width=0.75\textwidth]{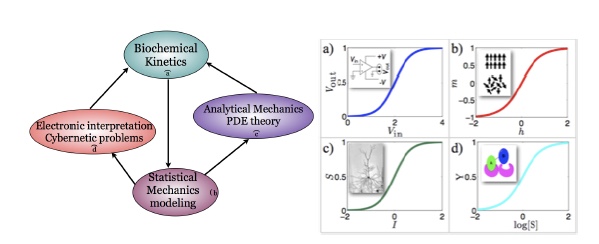}
\par\end{centering}
\caption{\textbf{Behavioral and formal analogies highlighted and exploited in this work}. Left panel: the whole logical procedure to understand information processing during chemical reactions is schematically shown.  (a) We deal with a problem in biochemistry. (b) Then we model it into a statistical mechanical framework, that (c) it is then solved via techniques of analytical mechanics and (d) whose results are finally interpreted in cybernetical/electronical terms. These findings can be further translated back into the biochemical scenario, whose modus operandi has now become transparent. Right panel:  Relevant behavioral analogies in the response of these saturable different systems lie at the basis of the structural equivalence we use and celebrated examples are reported.
(a) The sigmoidal shape of the transfer function of an operational amplifier, where the response is the output voltage while the stimulus is the input voltage.
(b) The sigmoidal shape of the self-consistency of a ferromagnet, where the response is the magnetization and the stimulus is the external magnetic field.
(c) The sigmoidal response of the activation function of a neuron where the output is the action potential voltage while input is conveyed by all the afferent electric synaptic currents.
(d) The sigmoidal shape of the saturation curve of a cooperative chemical reaction, where the output is the fraction of bound sites, while the input is the (log-)concentration of the ligand.}
\label{fig:ruocco}
\end{figure}

\begin{figure}[tb]
\noindent \begin{centering}
\includegraphics[width=0.65\textwidth]{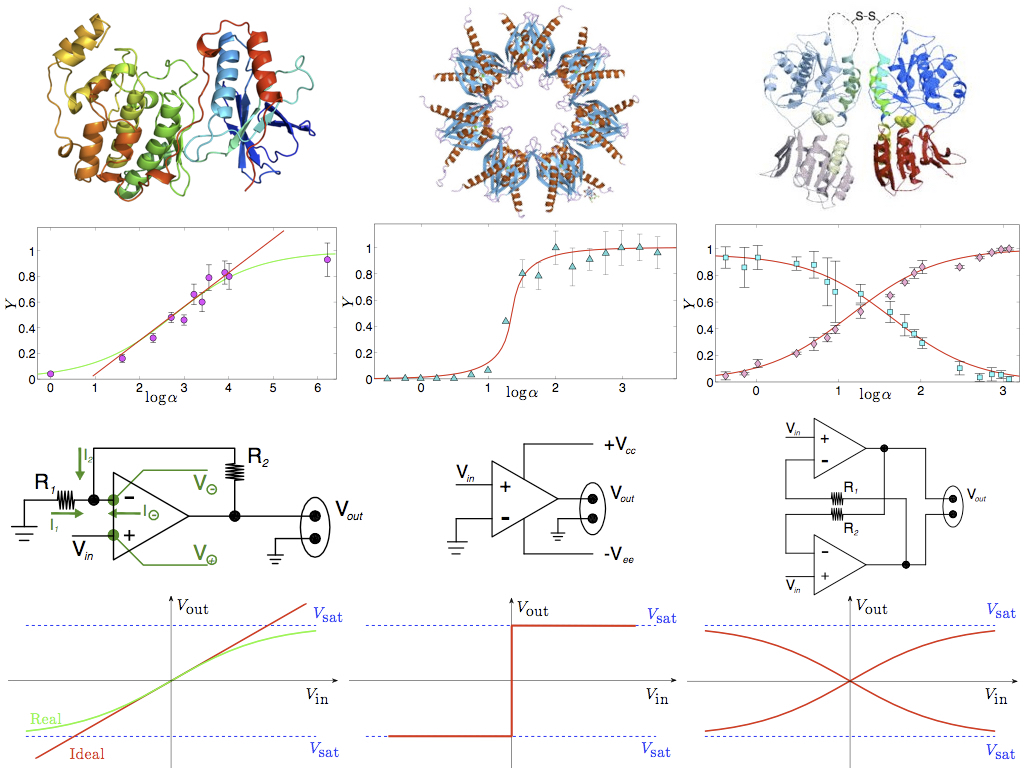}
\par\end{centering}
\caption{This figure summarizes the structural analogy between biochemical and electronic information processing. In the top-line three different proteins are shown: from left to right, the first (mitogen-activated protein kinase 14) obeys cooperative kinetics, the second (calmodulin dependent protein kinase 2) ultra-sensitive kinetics, while the last (synaptic glutamate  receptor) fulfills anti-cooperative kinetics.
In the second row the saturation curves for these proteins are shown: $\alpha$ is the concentration of the free ligand needed for their binding and $Y$ is the fraction of occupied binding sites. Symbols with the relative error-bars stand for real data taken from  \cite{solomatin,bradshaw,Suzuki-JBC2004} and lines are best fits performed through the analytical expression in Eq.~(\ref{trita}), where the best-fit coefficient $J$ corresponds to an ergodic ferromagnet (left), a low-temperature ferromagnet (center) and an anti-ferromagnet (right).
In the third row three circuits are shown. From left to right, an operational-amplifier, an analog-to-digital converter and a flip-flop, while in the fourth row their transfer functions are presented to highlight the behavioral analogy with the proteins. The latter shows the response of the system (output voltage) as a function of the input of the system (input voltage).
Following columns instead of rows, the first column is due to cooperative behavior, the second to ultrasensitive behavior and the last to anticooperative behavior.
\newline
The authors are grateful to Nature Publishing Group for the permission to reproduce this figure from \cite{ABBDU-SciRep2014}.}
\label{fig:figurone}
\end{figure}

Restricting to steady states, an alternative approach relies on statistical mechanics, as suggested by C.J. Thompson in his seminal work  \cite{thompson}. Indeed, statistical mechanics turns out to be particularly effective for the description of universal behaviors of a wide range of biological systems (from the extra-cellular level of neural \cite{hopfieldnature,amit,Agliari-PRL2012} or immune networks \cite{Kardar-PNAS2013,bialek-MEA,AABCT-JPA2013}, to the intracellular level of gene regulatory and protein networks \cite{stanley,bialek-MEA,shapiro}). Moreover, as recently observed in the case of large systems \cite{ABBDU-SciRep2014,AABDK-SciRep2015}, the statistical mechanical approach plays the role of a general stochastic framework that naturally highlights the structural and conceptual analogies between response functions in biochemical reaction kinetics and transfer functions in cybernetics (see Fig.s~\ref{fig:ruocco} and~\ref{fig:figurone}), thus tacitely working as a {\em translator} between these two worlds, that is crucial to show how information is handled by these biochemical systems.
\newline
However, the current research in biochemical information processing has been recently attributing particular importance to systems involving a relatively small number of units and this implies that the standard statistical mechanical picture, given in the thermodynamic limit (where the role of intrinsic noise can be suppressed \cite{PCK}), is not accurate. One of the goals of the present work is to extend the theoretical framework developed in \cite{ABBDU-SciRep2014,AABDK-SciRep2015} to allow for the description of systems of finite sizes.\\

Before proceeding it is worth stressing that the statistical mechanical description of chemical kinetics pursued here is based on the \emph{canonical ensemble} and it is accomplished at the \emph{mean-field} level. This is therefore clearly different  from the statistical mechanical description based on the gran-canonical ensemble introduced in the $70$'s \cite{GCE}. The advantage of the present formulation is that the use of a canonical framework allows us to establish structural bridges between the phenomenology of these bio-chemical systems on one side and the well-consolidated theory of information processing systems (i.e. cybernetics) on the other side, since the latter (inflected for instance in terms of neural networks and learning machines) is mainly developed within the canonical formalism \cite{amit}.
\newline
Along the same line we choose to adopt the mean-field perspective: admittedly, this implies a cost (as we give up a detailed description of the true architecture of the system under consideration and we deal with effective parameters to be properly renormalized), yet the reward lies in the possibility to directly compare the emerging response functions with transfer functions in cybernetics and therefore to understand how information is processed through  a given reaction. Further, trough direct calibration of the re-normalized key parameters over a small subset of data, the whole theory become of immediate experimental applicability.

\begin{figure}[tb]
\noindent \begin{centering}
\includegraphics[width=0.5\textwidth]{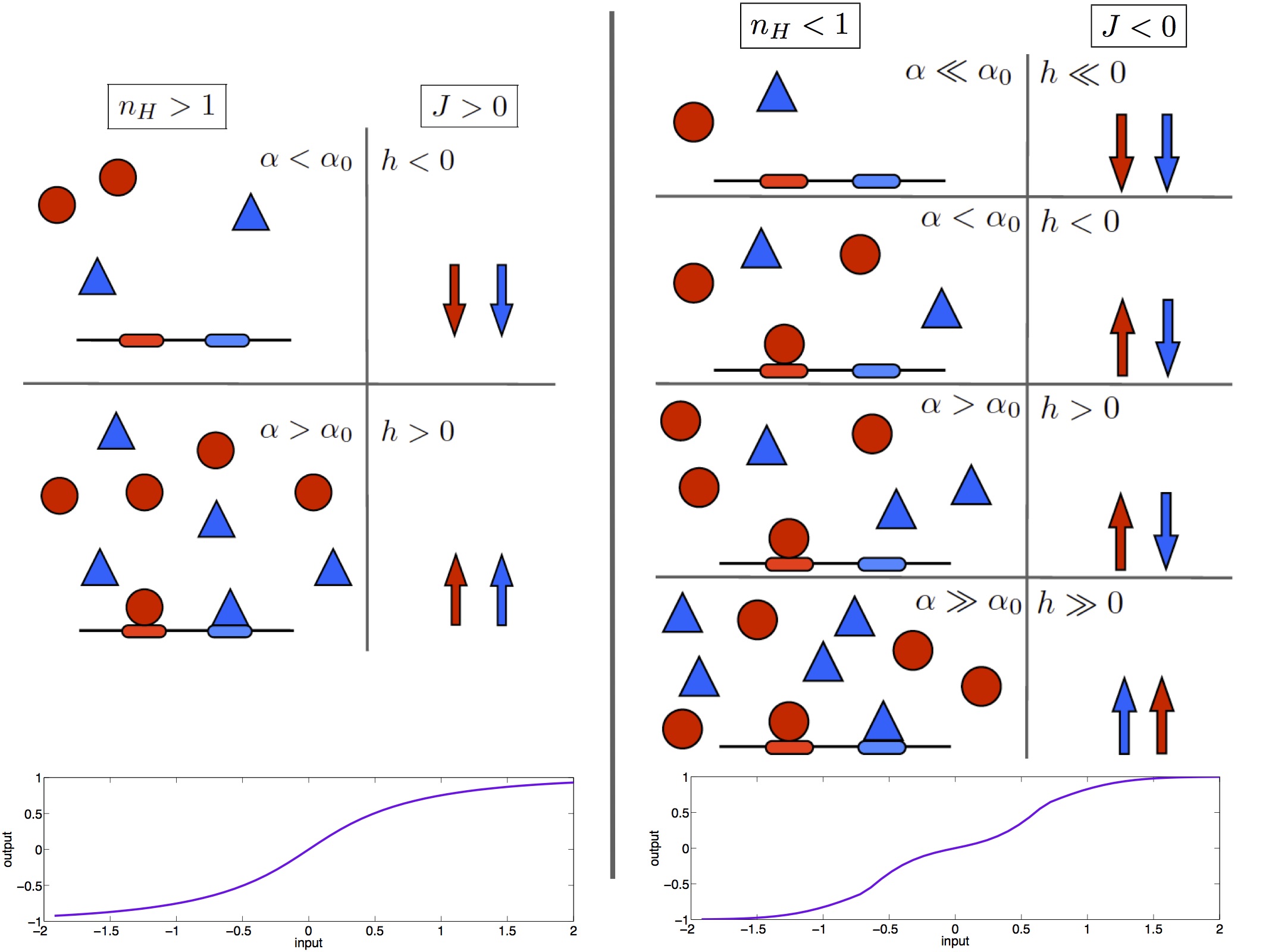}
\caption{Schematic representation of the processes considered. The cooperative case is shown in the left side, while the anti-cooperative case is shown in the right side. Two different binding sites, corresponding to two spins belonging to different parties, are shown in different colors. The cooperative case is characterized by a Hill coefficient $n_H$ larger than $1$, corresponding to a positive coupling $J$ between spins. The anti-cooperative case is characterized by a Hill coefficient $n_H$ smaller than $1$, corresponding to a negative coupling $J$ between spins. The input of the system is provided by the concentration of free ligand $\alpha$, corresponding to the magnetic field $h$  (more precisely, $h = 1/2 \log(\alpha / \alpha_0)$, where $\alpha_0$ is the half-saturation concentration, in such a way that a negative field is equivalent to a concentration smaller than $\alpha_0$ and vice versa). The output is provided by the saturation $Y$, corresponding to the magnetization $m$ (more precisely, $m = 2Y -1$, in such a way that the mean performed over the magnetizations of the two parties is equivalent to the sum of the saturations pertaining to the two binding sites minus $1$). The plots in the bottom represent the evolution of the output as the input is varied. In particular, the curves shown here are obtained as the solution of Eq.~(\ref{trita}) for $J=0.5$ and $J=-1.2$, respectively.}
\label{fig:cartoon}
\end{centering}
\end{figure}

Let us now present in more detail the statistical-mechanical framework adopted in this work and the results we obtain through this path. We consider large macromolecules (e.g., polymers, proteins, etc.), whose binding sites are dichotomic variables (Boolean logical operators to be thought of as Ising spins) that can be either empty or occupied by a ligand (e.g., a substrate): the log-concentration of free ligands plays the role of the external field in spin systems. The binding sites are not necessarily independent: in the special case where they are independent the emerging kinetics follows the so-called Michaelis--Menten law, and this is naturally captured by the lack of interaction between spins within the statistical mechanical route of formalization. On the contrary, if there is interaction, the kinetics can be cooperative (for positive values of the interaction), or anti-cooperative (for negative values of the interaction), and these behaviors are typically coupled to various names, e.g., Hill, Koshland, and Adair reactions, etc. (see e.g.  \cite{GCE}). More precisely, binding sites with identical structure/function (e.g., the four oxygen-capturing arms of the hemoglobin) are modeled as a unique spin system, where each spin feels the external field (the oxygen log-concentration in this example) and interacts imitatively with other spins; conversely, binding sites with different structure/function are modeled as a spin system fragmented into more parties, where spins belonging to different parties compete to bind the ligand by anti-imitative interactions  (e.g., for the two insulin-capture $\alpha$-subunits in the insulin receptor, the system describing the receptor is split into two main parties -i.e. {\em arms}-- that anti-cooperate).  This scheme naturally leads to a statistical mechanical scenario consisting in multi-partite spin systems (on which --in turn- there has also been recent interest in the Statistical Mechanical community \cite{bipartiti1,bipartiti2,bipartiti3,antonio,dmitry,Tantari-JSP2016}), with (positive or null) intra-party interactions and (positive or negative or null) inter-party interactions; we will focus on the two-party case, as schematized in Fig.~\ref{fig:cartoon}. The behavior of this system in the presence of an external magnetic field $h$ (i.e., the {\em input}) can be captured in terms of the average magnetization $m$ (i.e., the {\em output}), which keeps track of the underlying collective features among spins. From a kinetics perspective, the function $m(h)$ recovers the saturation function of a system of binding sites as the concentration of free ligands is tuned. The isotherm of the magnetization and the saturation function are just two different inflections of the same transfer function, namely of the same input-output relation.
\newline
In this paper the exact, explicit, solution of these magnetic systems (and of the reaction kinetics they code for too) is given also at finite volumes. In particular, we derive a set of linear multidimensional partial differential equations for the partition function of the bipartite spin model which are valid for any (either finite or infinite) number of spins $N$. The solution for the model is specified via a suitable initial datum. We show that in the thermodynamic limit $N \to \infty$ the free energy related to these systems fulfils a set of Hamilton--Jacobi type equations that is completely integrable by the characteristics method and we provide the explicit solutions in terms of partial magnetizations, which in turn satisfy a set of two coupled equations of state. Via this route, the way these reactions handle information processing becomes transparent; such equivalence is summarized in Fig. \ref{fig:figurone} by means of illustrative examples.
\newline
Once wrote the general theory, at first we successfully compare the expected behavior of our theoretical saturation function with experimental results for several examples of cooperative, anticooperative and ultrasensitive kinetics of large systems: we show that the key parameters of the theory match -always with remarkable accuracy- the standard empirical indicators (e.g., the coupling constant used in the Hamiltonian spin-like representation of the reaction recovers the Hill coefficient of standard kinetics literature).
\newline
Then, we proceed by considering {\it small systems} (i.e. away from the thermodynamic limit) and we show that the solution at finite sizes can still be obtained explicitly (by separation of variables once an Ansatz on the initial state is given): the framework obtained in this way is used to explore and address several phenomena recently highlighted in the experimental literature on small system's kinetics (e.g., we recover that systems devoiding of cooperativity can still display a cooperative-like behavior and, possibly, bistability phenomena due to stochastic effects, as for instance discussed in  \cite{Lipshtat-PRL2006}).

The paper is organized as follows. First, we provide a brief survey of the statistical mechanical formulation of reaction kinetics as proposed by Thompson in \cite{thompson} and later developed in \cite{ABBDU-SciRep2014,AABDK-SciRep2015}. Then, we derive differential identities for the partition function and the free energy related to these models and construct their solutions both in the thermodynamic limit and for small systems. Next, we discuss implications of our theory such as noise-induced cooperativity and bistability, signal amplification and noise suppression. Finally, we summarize and comment on our results and discuss further outlooks.

\section*{Results}
\subsection*{Standard chemical kinetics: statistical mechanical formalization and cybernetic interpretation}
Hereafter we review main concepts of chemical kinetics, statistical mechanics and cybernetics, which we will be needed in the following; we refer to classical textbooks for a more extensive treatment (see e.g., \cite{Mazza-2014,PCK,ellis,millman,wiener}).

\emph{\textbf{Chemical-kinetics framework.}} In many macromolecules (i.e. polymers, proteins, etc.) ligands bind in a non-independent way. In particular, if, upon a ligand binding, the probability of further binding (by other ligands) is enhanced, as for example in the paradigmatic case of hemoglobin \cite{thompson}, the system is said to exhibit positive cooperativity; viceversa, cooperativity is negative where further binding of ligands is inhibited \cite{koshland} as in the case of some insulin receptors \cite{insulina} (see Fig.~\ref{fig:cartoon} for a schematic representation).
Fundamental mechanisms underlying cooperativity depend, in general, on the microscopic details of the system under consideration. For instance, in the case of polymers, if two neighbor docking sites bind charged ions, the electrostatic attraction/repulsion may be responsible for positive/negative cooperativity.

In complete generality, let us consider a model where several identical hosting (macro)-molecules $P$ can bind overall $N$ identical small molecules $S$ (whose concentration is denoted by $[S]\equiv \alpha$) on their structure; calling $P_j$ the complex of a molecule $P$ with $j \in [0,N]$ molecules attached, at the chemical equilibrium we have
$$ \alpha+ [P_{j-1}] \rightleftharpoons [P_j].$$
For the sake of simplicity, let us consider the case $j=1$. The time evolution of the concentration $[P_0]$ of the unbound protein $P_0$ is governed by the equation
\be
\frac{d[P_0]}{dt} = - K_{+1}^{(1)} ~ [P_0] ~ \alpha +K_{-1}^{(1)} ~ [P_1],
\ee
where $K_{+1}^{(1)},  K_{-1}^{(1)}$ are, respectively, the forward and backward rate constants for the state $j=1$, and their ratio defines the {\em association constant} $K^{(1)} \equiv  K_{+1}^{(1)}  / K_{-1}^{(1)}$. In the steady state $d [P_{0}]/dt = 0$, we have
$$
 K^{(1)}=\frac{[P_1]}{[P_{0}]\alpha}.
$$
For generic $j>0$ we have
\begin{equation}
\label{Kiter}
 K^{(j)}=\frac{[P_j]}{[P_{j-1}]\alpha}.
\end{equation}
Therefore, in general, one can write $[P_1] = [P_0] K^{(1)} \alpha$, $[P_2] = [P_1] K^{(2)} \alpha = [P_0] K^{(1)} K^{(2)} \alpha^2$, and, by extension, $[P_j] = [P_0] ( \prod_{i=1}^j K^{(i)})  \alpha^j$.
In many practical situations, a direct measure of $[P_j]$ is not feasible. A convenient experimental observable is then given by the average number $\bar{S}$ of ligands that, at the equilibrium, are bound to the macromolecule(s) and this is given by
\be \label{eq:adair}
\bar{S} = \frac{[P_1]+2[P_2]+...+N[P_N]}{[P_0]+[P_1]+...+[P_N]} = \frac{K^{(1)} \alpha + 2 \, K^{(1)} K^{(2)} \alpha^2+...+ N \, K^{(1)} K^{(2)} K^{(3)} \dots K^{(N)} \alpha^N}{1+  K^{(1)}\alpha +  K^{(1)}K^{(2)} \alpha^2+...+ K^{(1)} K^{(2)} K^{(3)} \dots K^{(N)}\alpha^N}.
\ee
The last expression in the previous equation is the well-known Adair's equation \cite{PCK}, obtained by iterating the relation~(\ref{Kiter}).\\
In this work (as standard) we will focus on the normalized version of $\bar{S}$, called {\em saturation function} $Y$ and defined as
\be \label{eq:Ydef}
Y= \frac{\bar{S}}{N}.
\ee
Unless otherwise stated, $Y$ represents the expected fraction of occupied sites in the whole set of molecules $P$ under investigation, namely the global system's response (i.e. the output signal) to the stimulus provided by the ligand's concentration $\alpha$ (i.e. the input signal).
\newline
In a {\em non-cooperative} system, one expects independent and identical binding  with microscopic association constant $K$, and one can write $K^{(j)}= (N-j+1)\, K / j$. In this case, Adair's equation~(\ref{eq:adair}) for $\bar{S}$ and $Y$  reads as
\ba
\label{MM0}
\bar{S} &=& \frac{N K \alpha}{1+ K \alpha},\\
\label{MM}
Y &=& \frac{K \alpha }{1+K \alpha }.
\ea
The latter expression is the well-known Michaelis--Menten equation \cite{PCK}.

Clearly, the kinetics becomes far less trivial as soon as interactions among binding sites occur.
For the sake of simplicity, if we consider the limiting case where intermediate steps can be neglected, i.e.
$$  \alpha + [P_0] \rightleftharpoons [P_N],$$
we get that
\begin{eqnarray}
\bar{S} &=& \frac{N [P_N]}{[P_0]+[P_N]} = \frac{N K \alpha^N}{1+ \alpha^N}, \\
Y &=& \frac{K \alpha^N}{1+ \alpha^N}.
\end{eqnarray}
More generally, accounting also for some degree of sequentiality \cite{PCK}, one obtains the well-known Hill's equation
\be\label{collina}
Y = \frac{K \alpha^{n_H}}{1+\alpha^{n_H}},
\ee
where $n_H$, referred to as Hill coefficient, represents the effective number of interacting binding sites. We note that, by posing $n_H=1$, equation~(\ref{collina}) recovers the Michaelis--Menten law (\ref{MM}). In fact, in the non-cooperative case, the number of binding sites effectively interacting is just one (i.e., there are no true interactions). For $n_H > 1$ the kinetics is said to be {\em cooperative}, while for $n_H < 1$ it is said {\em anti-cooperative}.
If $n_H \gg 1$ the kinetics is said to be {\em ultra-sensitive}.
\newline
Given a set of experimental measures of the saturation function $Y(\alpha)$, $n_H$ is estimated as the slope of $\log[Y/(1-Y)]$ versus $\alpha$, measured at half-saturation (namely, when $Y=1/2$). \\

\emph{\textbf{The statistical mechanics formalization.}} Statistical mechanics can provide a convenient language to describe biochemical kinetics and frame it into a (noisy) logical scaffold (see \cite{ABBDU-SciRep2014,AABDK-SciRep2015} for details). In the present section we briefly review how main concept of mean-field cooperative statistical mechanics (i.e. ferromagnetism) apply to the scenario of cooperative reaction kinetics.
\newline
We start from the microscopic model described above, consisting of an ensemble of identical macromolecules, carrying overall $N$ binding sites labeled by $i=1,2,\dots$. Macromolecules are in a solution with smaller molecules (the substrate) and each binding site can accommodate only one molecule of the substrate. Notice that, following the mean-field approach, here we do not distinguish  binding sites belonging to the same molecule, but we are treating the whole set of binding sites, pertaining to the whole set of molecules, at once.

Let us associate to each binding site an Ising spin $\sigma$ where $\sigma_i=+1$ if the $i^\textrm{th}$ site is occupied and $\sigma_i=-1$ if it is empty. A particular configuration of the system is specified by the set $\{ \sigma \}$, through which all the classical observables can be introduced. For instance, the magnetization is defined as $m(\{\sigma\}) =  (\sum_{i=1}^N \sigma_i)/N$ and is directly related to the occupation number $\bar{S}$ of binding sites by
\be
\label{Sandm}
\bar{S}(\{\sigma\}) =  \sum_{i=1}^N \frac{1+\sigma_i}{2} =  \frac{N}{2} \left [1 + m (\{\sigma\}) \right ],
\ee
in such a way that the saturation function (in terms of the magnetization) reads as
\be
\label{Yandm}
Y(\{\sigma\}) = \frac{\bar{S}(\{\sigma\})}{N} = \frac{1 + m (\{\sigma\}) }{2}.
\ee
For simplicity, we first consider a system which does not exhibit collective behavior, that is, no interaction between binding sites is present and only the interaction between the binding sites and the ligand occours. The external field in  statistical mechanics, namely a scalar parameter $h$, is introduced as the log-concentration $\log(\alpha)$ of free-ligand molecules \cite{thompson}. As there are no spin-spin interactions, this system is mapped into a system of free spins $\sigma$ thermalyzing in the presence of a magnetic field $h$ and whose energy function (or {\em Hamiltonian}) is given by
\be\label{MM1}
H(\{ \sigma \},h)=-h\sum_{i=1}^N \sigma_i.
\ee
Note that $h$ is assumed to be independent of the site index, meaning that binding sites interact uniformly and independently with the ligands; this also means that the time-scale for diffusion is fast with respect to the time scale for thermalization.

Once the Hamiltonian representing the model under investigation is defined, we can apply the standard statistical mechanics machinery, namely the Maxwell--Boltzmann probability distribution $P(\{\sigma\})$ associated to a generic system state $\{ \sigma \}$
$$
P(\{\sigma\}) = \frac{e^{-\beta H}}{Z}, \ \ \  \ \  Z=\sum_{\{ \sigma \} }^{2^N} e^{-\beta H}
$$
where $Z$ is the normalization, called {\em partition function} and $\beta$ tunes the level of noise in the system, in such a way that for $\beta \to 0$ the probability measure becomes flat and every state assumes the same chance to happen, while for $\beta \to \infty$ the system collapses in configurations corresponding to the minima of the Hamiltonian.
\newline
Throughout the paper, given a generic function of the spins $f(\sigma)$, the bracket averages $\langle f(\sigma) \rangle$ represent the averages over the Maxwell--Boltzmann probability distribution.
\newline
An explicit evaluation of the partition function $Z$ allows for a direct measurement of the free energy $F$ related to the model encoded by $H$, that is, $F = (1/N) \ln Z$ (notice that, in our derivation, just for mathematical convenience we work with the negative free energy $\tilde{F} = - (1/N) \ln Z$; hence, $\max(F)$ just corresponds to $\min(\tilde{F})$). In general, $F$ depends on the system configuration $\{\sigma\}$ and on the system parameters, namely the external field, the level of noise, and the possible coupling between spins. Once these parameters are set, the maximization of $F$ provides the related equilibrium state(s). This is because the free energy $F$ is nothing but $F = - \beta \mathcal{U} + \mathcal{S}$, where $\mathcal{U} = \langle H \rangle$ is the internal energy of the system (i.e. the Maxwell-Boltzmann average of the Hamiltonian at given noise $\beta$) and $\mathcal{S}= -\langle \ln P(\{ \sigma \}) \rangle$ is the entropy, thus, maximizing $F$ implies looking for the minimum of $\mathcal{U}$ and, simultaneously, the maximum of $\mathcal{S}$ (at given $\beta$).
\newline
In particular, for the system described by the Hamiltonian (\ref{MM1}), the energy is minimized by those configurations where spins are aligned with the external field. Thus, in the absence of any source of noise, the system relaxes to a state where $\av{m}=1$ (i.e., $\sigma_i = +1, \forall i$) if $h > 0$, or to a state where $\av{m}=-1$ (i.e., $\sigma_i = -1, \forall i$) if $h < 0$. In the presence of noise the equilibrium state requires, beyond energy minimization, also entropy maximization:
setting for simplicity $\beta =1$ (without loss of generality since this is equivalent to rescale $h \rightarrow h \beta$) one can find that the equilibrium state corresponds to $\av{m} = \tanh (h)$ (see e.g., \cite{ellis}).
\newline
Actually, one could reach the same result without relying on statistical mechanics, as explained hereafter. First, we need to relate $h$ to the ligand concentration $\alpha$.
As mentioned above, the energy is minimized by those configurations where spins are aligned with the external field: if $h>0$ molecules tend to bind to diminish energy, while if $h<0$ bound molecules tend to leave occupied sites; this suggests that $h$ plays the role of the chemical potential for the binding of the ligand molecules on the docking sites. Moreover, as observed in \cite{thompson,ABBDU-SciRep2014,AABDK-SciRep2015}, the chemical potential can be expressed as the logarithm of the concentration of the substrate, upon proper normalization, namely
\be \label{eq:ha}
h = \frac{1}{2} \log \left ( \frac{\alpha}{\alpha_0} \right),
\ee
where $\alpha_0$ stands for the value of ligand concentration such that binding sites have the same probability of being occupied or unoccupied.
\newline
Crucially, under the mean-field approximation (and this is the only case, apart from linear chain models which, still, do not exhibit  phase transitions and are of moderate interest), the probability $P(\{ \sigma \})$ of the configuration $\{ \sigma\}$ can be factorized as $P(\{ \sigma \})=\prod_{i=1}^N P(\sigma_i)$. One can therefore focus on the single-spin probability and, following \cite{thompson}, state that $P(+1)$ is proportional to the concentration $\alpha$ and  that $P(-1)$ is proportional to the inverse of the concentration $\alpha$, that is, $P(+1)=C e^{+h}$ and $P(-1)=C e^{-h}$, where $C$ is fixed in such a way that $P(+1)+P(-1) = 1$, i.e. $C= 1/(2 \cosh(h))$. Then, the expression $P(\sigma_i) = \exp( h \sigma_i) / [2\cosh(h)]$ implies that
\[
\langle m(h) \rangle = \sum_{\{ \sigma \}} \prod_{i=1}^N P(\sigma_i) \sigma_i = \tanh(h).
\]
Exploiting this result the average saturation function (see Eq.~(\ref{Yandm})) reads as
\be \label{eq:Y}
\av{Y(\alpha)}=\frac12 [1+ \tanh (h)].
\ee
Using Eq.~(\ref{eq:ha}),  and recalling that $\tanh (x) = [\exp(2x) - 1]/[\exp(2x)+1]$, it is immediate to check that Eq.~(\ref{eq:Y}) coincides with the Michaelis--Menten equation~(\ref{MM}) with the particular choice $K=1/ \alpha_0$. This is perfectly consistent with the underlying assumption of independent binding sites (i.e. no couplings among spins) tacitly made when we defined the system under study, via the Hamiltonian (\ref{MM1}).

Let us now generalize this scenario by introducing the simplest possible two-body interaction given by the Hamiltonian
\be\label{MMCurie}
H(\{ \sigma \},h,J)=- \frac{J}{2 N}  \sum_{i,j}^{N,N}  \sigma_i \sigma_j -h\sum_{i=1}^N \sigma_i.
\ee
The Hamiltonian~(\ref{MMCurie}) represents the well-known Curie--Weiss theory of ferromagnetism \cite{ellis}. The first sum in the right hand side of~(\ref{MMCurie}) accounts for all possible $N(N-1)/2$ interacting pairs of spins and the coupling is homogeneous as the constant $J$ is the same for all pairs. If $J>0$ the configurations where spins are aligned are more favored thus this choice naturally leads to a theory for cooperative kinetics, while  $J<0$ favors the configurations where spins compete and are misaligned thus working for the anti-cooperative case. Clearly, models with different values of $J$ represent different chemical systems.

As well known, the condition of minimization for the Curie--Weiss free energy in the thermodynamic limit $N \to \infty$, yields the so-called self-consistency equation (see e.g., \cite{ellis})
\begin{equation}\label{self}
\langle m \rangle = \tanh \left[ \beta J  \langle m \rangle + \beta h   \right].
\end{equation}
Recalling the mapping~(\ref{Yandm}), and reabsorbing $\beta$ by the rescaling $\beta J \to J$ and $\beta h \to h$, the previous equation translates into the reaction kinetics vocabulary as
\begin{equation}\label{trita}
Y(\alpha) = \frac{1}{2} \left \{ 1 + \tanh \left[ J(2 Y -1) + \frac{1}{2} \log \left( \frac{\alpha}{\alpha_0} \right) \right] \right \}.
\end{equation}
Eq.~(\ref{self}) implies that, at low noise levels, the Curie--Weiss model exhibits an abrupt change in the magnetization as a function of $h$.
More precisely, a second order phase transition occurs at $h=0$ (i.e., $\alpha = \alpha_0$) so that $m$, and then $Y$, is continuous while its derivative diverges as a function of $J$ at the critical value $J_c = 1$. The phase transition is first order if, at fixed $J > J_c$, $m$ (and then $Y$), is viewed as a function of $h$. In this case, for $J> J_{c}$, a jump occurs at $h=0$ and the reaction is {\em ultra-sensitive} (its transfer function mirrors that of an ON/OFF switch).

As a robustness check, we verify that, in the absence of interaction, the model is consistent with  the classical Michaelis--Menten kinetics. Indeed, Eq.~(\ref{trita}) can be written as
\be\label{selfcoop}
Y(\alpha;J)= \frac{\alpha e^{2J (2Y-1)}}{\alpha_0+\alpha e^{2J(2Y-1)}},
\ee
and, setting $J=0$, the equation above clearly reduces to Eq.~(\ref{MM}) with $K=\alpha_0^{-1}$. In this case, no phase transitions occur and the model turns out to be not suitable to describe complex biochemical reactions.

As anticipated, in order to provide a quantitative information about how a system actually departs from the simplest Michaelis--Menten framework, the so-called Hill coefficient is introduced, defined as the slope of $\log[Y/(1-Y)]$ versus $\alpha$, which can be recast as
\be\label{hill}
n_H =  \frac{1}{Y(1-Y)} \left . \frac{\partial Y}{\partial \alpha} \right |_{Y=1/2}= \frac{1}{1-J},
\ee
where the last expression was obtained using Eq.~(\ref{trita}). It is straightforward to check that the Hill coefficient for the Michaelis--Menten model, obtained for $J=0$, is  $n_{H} = 1$ as expected. On the other hand, a positive value of $J$ implies a positive cooperativity and the closer $J$ is to the critical value $J_c= 1$ (from below) and the stronger the cooperativity exhibited by the system; values of $J$ larger than $1$ correspond to discontinuous saturations (i.e. systems showing ultra-sensitive responses).

We stress that the relation (\ref{hill}) provides a crucial link between the experimentally accessible quantity $n_{H}$ and the theoretical parameter $J$, namely the coupling strength of the underlying mean-field spin description.

We finally observe that small-coupling expansions of the right hand side of (\ref{trita}), or, equivalently, of (\ref{selfcoop}), lead to suitable polynomial approximations for the saturation function to be possibly compared with formulas obtained through classical  (usually ODE-based) phenomenological approaches in chemical kinetics \cite{GCE}. For example, the first order expansion of the expression (\ref{selfcoop}) around $J=0$ gives
\begin{equation}\label{eq:Ycoop}
Y(\alpha) \approx \frac{(1-J) \alpha + \alpha^2}{1+2(1-J)\alpha+ \alpha^2 },
\end{equation}
which is equivalent  to Adair's equation~(\ref{eq:adair}) provided that $J= (1-\sqrt{{K^{(1)}}^3 {K^{(2)}}}/2)$ along with the rescaling $\alpha \rightarrow \alpha/\sqrt{K^{(1)} K^{(2)}}$.

\emph{\textbf{The cybernetic interpretation.}}
The cybernetic interpretation of chemical kinetics, extensively treated in \cite{ABBDU-SciRep2014,AABDK-SciRep2015}, is based on the behavioral analogy between the saturation curves (or binding isotherms) in chemical kinetics, the self-consistencies (i.e. state-equations) in statistical mechanics and the transfer functions in electronics, see Fig.~\ref{fig:figurone}. Similarly to saturation curves, self-consistency functions and transfer functions are nothing but relations between an input (the ligand concentration $\alpha$ in chemical kinetics, the magnetic field $h$ in statistical mechanics and the input voltage $V_{\textrm{in}}$  in electronics) and an output (the saturation function $Y$ in chemical kinetics, the expected magnetization $\langle m\rangle$ in statistical mechanics and  the output voltage $V_{\textrm{out}}$ in electronics).

As an illustrative example, let us consider the paradigmatic operational amplifier. In a regime of small input voltage $V_{\textrm{in}}$, the output voltage $V_{\textrm{out}}$,  is described by the following {\em transfer function}
\be \label{coop_V}
V_{\textrm{out}}= G \, V_{\textrm{in}} = (1+R_2) V_{\textrm{in}},
\ee
where $G=(1+R_2)$ is referred to as {\em gain} of the amplifier and $R_2$ is the feed-back resistor (allowing for true amplification as, if $R_2=0$,  then $G=1$), as shown in Fig.~\ref{fig:figurone} (left column). Similarities between this response function and the saturation curve in chemical kinetics can be clarified, using the statistical mechanics formalism, by comparing the mathematical structure of the response of these systems, namely comparing $V_{\textrm{out}}$ in~(\ref{coop_V}) with the average magnetization $\av{m}$ of a ferromagnet in the linear regime  [where $x=(J \av{m} + h)$ is small such that $\tanh(x)\sim x$]:
\be \label{coop_J}
\langle m \rangle = \tanh(J \langle m \rangle + h) \approx   J  \langle m \rangle  + h \Rightarrow \langle m \rangle \sim (1- J)^{-1} h.
\ee
Observing that, for $J<J_c \equiv 1$, the Hill coefficient~(\ref{hill}) can be approximated as $n_H = 1/(1-J) \sim (1+J)$, we can write
\begin{eqnarray}\label{compare}
V_{\textrm{out}} &=& (1+R_2) V_{\textrm{in}},\\
\av{m} &=& (1+J) h.
\end{eqnarray}
The conceptual term-to-term identification between inputs and outputs of the above transfer functions suggests that the Hill coefficient can be interpreted as the gain factor for the reaction (and we can already see why cooperativity is needed to amplify biochemical signals, as we require $J>0$). Let us also note that Hill coefficients are typically of order $\sim 10$ or less, and this contributes to explain why amplification of biochemical circuits is rather low \cite{winfree1} if compared to its electronic counterpart, where the gain can be of several orders of magnitude \cite{millman}.

We emphasize that this behavioral analogy between biochemical kinetics, statistical mechanics and electronics goes far beyond the linear regime exploited above because all the related response functions (i.e. binding isotherms, self-consistencies and transfer functions), far away from the linear regime, display intrinsic {\em saturation effects} (see Fig.~\ref{fig:ruocco}. right panel): roughly speaking, if all the macromolecular binding sites are occupied, no matter how further we increase the ligand concentration, the system will not vary its response that is already maximal. The same applies to spin systems as, once all the spins are aligned with the magnetic field, a further increase in the latter can not produce any shift in the system's response and the same holds for operational amplifiers too as, once the output voltage reaches the collector tension, any further increase in the input voltage will no longer produce a variation in the response.
\newline
Moreover, apart from the operational amplifier, also other devices naturally fit the picture provided for describing biochemical reactions in different regimes. For example, if $J > J_c$ (i.e., $n_H \gg 1$ and $G \gg 1$) the expected magnetization $m(h)$ as well as the saturation function $Y(\alpha)$ develop a discontinuity at $h=0$ and $\alpha=\alpha_0$ respectively (and we refer to ultra-sensitive kinetics in biochemistry and first-order phase transitions in statistical mechanics). The corresponding limit for the operational amplifier is the analog-to-digital converter (ADCs) at $V_{\textrm{in}}=0$ (see Fig.~\ref{fig:figurone}, center column).
\newline
Another example is given by the simplest bistable flip-flop, constituted by two saturable operational amplifiers reciprocally inhibiting, in such a way that the output of one of the two amplifiers is used as the inverted input of the other amplifier: this is the simplest 1-bit memory device as it is possible to assign a logical 0 (or 1) to one state and the other logical 1 (or 0) to the other state. As the two operational amplifiers (i.e. the two parties) interact   inhibiting reciprocally, the translation of this circuit into a chemical circuit will require anti-cooperativity among the two parties, thus will be naturally accounted when anti-cooperativity is at work (see Fig.~\ref{fig:figurone}, right column).
\newline
When dealing with more complex reactions, such as reactions involving several components and several steps, our approach allows to recognize the various basic 'devices' at work and to build the whole circuitry in a cascade fashion so to figure out how the equivalent 'electronic circuit' effectively processes information as the reaction goes by.
In particular, a suitable combination of the elementary bricks described above leads to the construction of (bio-)logic chemical gates, as discussed in \cite{AABDK-SciRep2015}.
\newline
Here, rather than exploring further that route, we keep the focus on the outlined fundamental bricks and analyze their behavior away from the thermodynamic limit.

\section*{The {\em mechanical} formulation of chemical kinetics}
The analogies between the saturable systems described in the previous section have been developed in \cite{ABBDU-SciRep2014} restricting to the thermodynamic limit. This is a plausible regime for experiments involving extensive solutions of reactants and, indeed, the thermodynamic limit underlies also standard approaches in early modeling (e.g., classical chemical-reaction kinetics) \cite{vankampen,levitzki,koshland,Tian}. However, thanks to novel experimental breakthroughs, finally a number of recent studies involves only small numbers of molecules thus questioning the validity of any description in terms of such a large scale limit \cite{TDL}. This broad class of novel experiments includes toggle switches \cite{Warren,Kauffman-JIntBiol2006,Lipshtat-PRL2006}, stochastic bistability \cite{kardar-bistability,Leibler-Nature2000,transcriptional-scenario}, reactions with noise-induced cooperativity \cite{McKane,zhang2,Bialek-PRL2008} and the whole quorum sensing \cite{Kardar-PNAS2013,chemical-master-equation,quorum-bacteria,Agliari-JTB2015}, just to cite a few: interestingly, these novel experiments in small systems have highlighted such complex behaviors which can not be recovered from theories based on the thermodynamic limit.
Scope of the current section is thus to extend the mapping described in the previous section in order to include small systems as well. This will be obtained in two steps. First, we need to frame the whole statistical mechanical  treatment of chemical kinetics into the mathematical scaffold of classical mechanics; the latter, extensively relying on non linear PDEs techniques, allows for an optimal mathematical control of these systems at finite sizes \cite{noi-prsa1,noi-prsa2,noi-AoP1,noi-AoP2,denittis}. Then, through this route, we extend the statistical mechanical treatment of reaction kinetics even to the case of finite systems and solve for the latter. Finally, we check the overall robustness of our theoretical predictions by recovering these novel outlined  phenomena.
\newline
In the following we first present the general mapping, then we handle the simpler case $N \rightarrow \infty$ to check that the known limit is properly recovered, and finally we address the finite-$N$ case in detail. We emphasize that, in both the regimes, the model is exactly solvable as a consequence of the complete integrability of the PDEs derived for the partition function and the (related) system's free energy.

\emph{\textbf{The differential identities for the partition function.}} 
Let us consider the statistical mechanical formulation of a system where binding sites are of two kinds and evaluate the exact partition function and the related free energy in the thermodynamic limit ($N \to \infty$) and in the case of finite volumes ($N \ll  \infty$). The present theory can be straightforwardly extended to account for an arbitrary number of different binding sites \cite{bipartiti1,bipartiti2,bipartiti3} and to address chain reactions (whose implementation can be helpful in several biochemical  information processing systems \cite{winfree1,winfree2}).

The general model we consider is described by a Hamiltonian of the form
\be
\label{bizione}
H_N = - \left  [\frac{J_{AB}}{N} \sum_{i=1}^{N_{A}} \sum_{j=1}^{N_{B}}  \sigma_{i} \tau_{j} + \frac{J_A}{N} \sum_{i=1}^{N_{A}} \sigma_{i} \sigma_{j} + \frac{J_B}{N} \sum_{i=1}^{N_{B}} \sum_{j=1}^{N_{B}} \tau_{i} \tau_{j} + h_A \sum_{i=1}^{N_{A}}  \sigma_{i} + h_B \sum_{i=1}^{N_{B}}  \tau_{i} \right],
\ee
where $\sigma$ and $\tau$ are Ising spins associated to the binding sites of type $A$ and $B$ respectively, $J_{AB}$, $J_{AA}$, $J_{BB}$ are the pairwise-coupling constants between sites (of type $A$ and type $B$, both of type $A$, both of type $B$, respectively), and the external fields $h_A$ and $h_B$ correspond to the chemical potentials associated to the party $A$ and to the party $B$, respectively.
The overall number of sites is $N =N_{A} + N_{B}$, in such a way that, setting $\gamma = N_{A}/N$, we have
\[
N_{A} = \gamma N, \qquad N_{B} = (1-\gamma) N.
\]
The relative magnetizations are defined as
\begin{align}
& m_A = \frac{1}{N_A}\sum_{i=1}^{N_A} \sigma_i, \qquad m_B = \frac{1}{N_B}\sum_{j=1}^{N_B} \tau_j.
\end{align}
The partition function for the system can then be written as
\be
\label{Zdef}
Z = \sum_{\{ \sigma \}}  \sum_{\{ \tau \}} e^{-\beta H} =\sum_{\{ \sigma \}}  \sum_{\{ \tau \}} \exp \left \{  N  \left [ t_{A} m_{A}^{2} + t_{AB} m_{A} m_{B} + t_{B}^{2} m_{B}^{2}+ x_A m_{A} + x_B m_{B} \right ] \right \}
\ee
where we have defined
\[
t_{A} =\gamma^{2} J_{AA} \beta,  \qquad t_{AB} = \gamma (1-\gamma) J_{AB} \beta, \qquad t_{B} = (1-\gamma)^{2} J_{BB} \beta, \qquad x_A = \gamma h_A \beta, \qquad x_B = (1-\gamma) h_B \beta.
\]
By direct differentiation one can immediately verify that  $Z$ satisfies the following set of compatible PDEs
\begin{gather}
\label{Zpdes}
\begin{aligned}
&\der{Z}{t_{A}} = \frac{1}{N} \dersec{Z}{x_A}, \\
&\der{Z}{t_{B}} = \frac{1}{N} \dersec{Z}{x_B}, \\
&\der{Z}{t_{AB}} = \frac{1}{N} \dermixd{Z}{x_A}{x_B}.
\end{aligned}
\end{gather}
Based on formula~(\ref{Zdef}), it can be proven  that the solution $Z(x_A,x_B,t_{A},t_{AB},t_{B})$ must satisfy the initial condition
\be
\label{Zpdeinit}
Z_{0} = Z(x_A,x_B,0,0, 0) = \left [2 \cosh \left( \frac{x_{A}}{\gamma} \right) \right ]^{\gamma N} \left [2 \cosh \left( \frac{x_{B}}{1-\gamma} \right) \right ]^{(1-\gamma) N}.
\ee
The free energy $F$ of the system, defined as $F = \frac{1}{N} \log Z$, consequently fulfils  the set of equations
\begin{gather}
\label{ABalphafull}
\begin{aligned}
& \der{F}{t_{A}} = \left( \der{F}{x_A} \right)^{2} + \frac{1}{N} \dersec{F}{x_A} \\
& \der{F}{t_{B}} = \left( \der{F}{x_{B}} \right)^{2} + \frac{1}{N} \dersec{F}{x_{B}} \\
& \der{F}{t_{AB}} = \der{F}{x_{A}} \der{F}{x_{B}} + \frac{1}{N} \dermixd{F}{x_{A}}{x_{B}},
\end{aligned}
\end{gather}
with initial condition $F_{0}(x_{A},x_{B}) = F(x_{A},x_{B},t_{A} =t_{B} =t_{AB} =0)$, where
\be
\label{alphainit}
F_{0}(x_{A},x_{B}) = \log 2 + \gamma \log \cosh \left(\frac{x_{A}}{\gamma} \right) + (1-\gamma)  \log \cosh \left(\frac{x_{B}}{1-\gamma} \right).
\ee

\emph{\textbf{The thermodynamic limit.}} 
In this section we evaluate the free energy and the equations of state for the bi-partite system (\ref{bizione}) in the thermodynamic limit $N\to \infty$, where the limit is taken keeping the ratio $\gamma = N_{A}/N$constant (that is, none of the two parties is negligible w.r.t. the other).  Hence, neglecting $O(1/N)$ terms in equations~(\ref{ABalphafull}), we have the  following system of PDEs
\begin{align}
\label{ABalpha}
& \der{F}{t_{A}} = \left( \der{F}{x_{A}} \right)^{2}, \qquad  \der{F}{t_{B}} = \left( \der{F}{x_{B}} \right)^{2}, \qquad \der{F}{t_{AB}} = \der{F}{x_{A}} \der{F}{x_{B}}.
\end{align}
The equations above are expected to provide an accurate description of the thermodynamic solution within regions of thermodynamic variables $\{x_{A}, x_{B}, t_{A},t_{AB},t_{B} \}$ such that the second order derivatives in equations~(\ref{ABalphafull}) are bounded.
We observe that the system of equations~(\ref{ABalpha}) is a completely integrable set of Hamilton--Jacobi type equations and the general solution can be calculated via the method of characteristics (see e.g.~\cite{Courant}). Hence we find that the general solution can be written as follows
\be \label{eq:FF}
F = \av{m_{\A}} x_{A}  + \av{m_{\B}} x_{B} + \av{m_{A}}^{2} t_{\A} +  \av{m_{\A}} \av{m_{\B}} t_{\A \B}+ \av{m_{\B}}^{2} t_{\B} - w(\av{m_{\A}},\av{m_\B})
\ee
where $\av{m_{\A}}$ and $\av{m_{\B}}$ are functions of the thermodynamics variables $x_{A},x_{B},t_{\A},t_{\A \B},t_{\B}$ defined by
\[
\av{m_{\A}} = \frac{\sum_{\{ \sigma \}}  \sum_{\{ \tau \}} m_{\A} e^{-\beta H}}{Z}, \qquad  \av{m_{\B}} = \frac{\sum_{\{ \sigma \}}  \sum_{\{ \tau \}} m_{\B} e^{-\beta H}}{Z}
\]
which can also be written as
\[
\langle m_{A} \rangle = \der{F}{x_{A}}, \qquad \langle m_{B} \rangle = \der{F}{x_{B}}.
\]
The functions $\av{m_{\A,\B}}(x_{A},x_{B},t_{\A},t_{\A \B},t_{\B})$ are obtained extremizing the free energy $F(x_{A},x_{B},t_{A},t_{AB},t_{B})$, i.e.
$$
\der{F}{\av{m_{A}}} = 0 \qquad \mbox{and} \qquad \der{F}{\av{m_{B}}} = 0,
$$
or, equivalently,
\begin{gather}
\label{bicrit}
\begin{aligned}
& x_A + 2 \av{m_{A}} t_{A} + \av{m_{B}} t_{AB} - \der{w}{\av{m_A}}  = 0,\\
&  x_B + 2 \av{m_{B}} t_{B} + \av{m_{A}} t_{AB} - \der{w}{\av{m_B}} = 0,
\end{aligned}
\end{gather}
where the function $w(\av{m_{A}},\av{m_{B}})$ is uniquely fixed via the initial condition on $\av{m_A}$ and $\av{m_B}$
\begin{align*}
&\av{m_A}(x_{A},x_{B},t_{A}=t_{B}=t_{AB} = 0) = \der{F_{0}}{x_{A}} = \tanh \left ( \frac{x_{A}}{\gamma}\right), \\
&\av{m_B}(x_{A},x_{B},t_{A}=t_{B}=t_{AB} = 0) = \der{F_{0}}{x_{B}} = \tanh \left ( \frac{x_{B}}{1-\gamma}\right),
\end{align*}
thus giving
\begin{gather}
\label{biw}
\begin{aligned}
w =& \gamma \left [ \left( \frac{1+\av{m_{A}}}{2} \right) \log \left(\frac{1+\av{m_{A}}}{2}  \right)  +\left( \frac{1-\av{m_{A}}}{2} \right) \log \left(\frac{1-\av{m_{A}}}{2}  \right)  \right]+ \\
&(1-\gamma) \left [ \left( \frac{1+\av{m_{B}}}{2} \right) \log \left(\frac{1+\av{m_{B}}}{2}  \right)  +\left( \frac{1-\av{m_{B}}}{2} \right) \log \left(\frac{1-\av{m_{B}}}{2}  \right)  \right].
\end{aligned}
\end{gather}

By evaluating equations~(\ref{bicrit}) for the function $w$ given in~(\ref{biw}), we obtain the self-consistency equations in the form
\begin{gather}
\label{selfie}
\begin{aligned}
\av{m_{A}} =& \tanh \left(\frac{x_{A} +  2 \av{m_{A}} t_{A} + \av{m_{B}} t_{AB} }{\gamma} \right) \\
\av{m_{B}} =& \tanh \left(\frac{x_{B} +  2 \av{m_{B}} t_{B} + \av{m_{A}} t_{AB} }{1-\gamma} \right).
\end{aligned}
\end{gather}
We observe that the equations of state~(\ref{bicrit}) can be interpreted as the hodograph solution (see e.g.~\cite{Tsarev}) of the following set of $(1+1)$-dimensional equations of hydrodynamic type for the partial magnetizations $\langle m_{A} \rangle$ and $\langle m_{B} \rangle$
\begin{align}
&\der{\av{m_{A}}}{t_{j}} = 2 \av{m_{j}} \der{\av{m_{A}}}{x_{j}}&  &\der{\av{m_{B}}}{t_{j}} = 2 \av{m_{j}} \der{\av{m_{B}}}{x_{j}}&&  \\
&\der{\av{m_{j}}}{t_{AB}} = \der{}{x_{j}} \left(\av{m_{A}} \av{m_{B}} \right)&
&\der{\av{m_{A}}}{x_{B}} = \der{\av{m_{B}}}{x_{A}}&& j = A,B.
\end{align}
The set of equations above is completely integrable as directly follows from the equations~(\ref{ABalpha}) with
\[
\av{m_{A}} = \der{F}{x_{A}}, \qquad \av{m_{B}} = \der{F}{x_{B}}.
\]
By differentiating equations~(\ref{selfie}), where magnetizations $\av{m_{A}}$ and $\av{m_{B}}$ explicitly depend on $x$ and $t$ variables, one can show that all first and second derivatives
$$
\der{\av{m_{A}}}{x_{A}}, \quad \der{\av{m_{B}}}{x_{B}}, \quad \dersec{\av{m_{A}}}{x_{A}}, \quad \dersec{\av{m_{B}}}{x_{B}}, \quad \dermixd{\av{m_{A}}}{x_{A}}{x_{B}}, \quad \dermixd{\av{m_{B}}}{x_{A}}{x_{B}},
$$
develop a gradient catastrophe on the hyper-surface
\begin{equation}
\label{hypers}
\gamma (1 - \gamma  ) + 2 (1 - \gamma) t_{A} (\av{m_{A}}^{2} -1) + 2 \gamma t_{B} (\av{m_{B}}^{2} -1) + (t_{AB}^{2} - 4 t_{A} t_{B}) (\av{m_{A}}^{2} + \av{m_{B}}^{2} - \av{m_{A}}^{2} \av{m_{B}}^{2} - 1) = 0.
\end{equation}
The solutions $\left(t^{(c)}_{A},t^{(c)}_{B},t^{(c)}_{AB},m^{(c)}_{A}, m^{(c)}_{B} \right)$ to the algebraic equation~(\ref{hypers}) are singular points on the space of coupling constants $\left(x_{A},x_{B},t_{A},t_{B},t_{AB} \right)$ where partial magnetizations develop a classical shock. Such shocks, as they are known in hyperbolic wave theory, are associated to  phase transitions in statistical mechanics \cite{bipartiti2,noi-prsa1,noi-AoP1,noi-AoP2,denittis}.
The equation~(\ref{hypers}) and the self-consistency equations~(\ref{selfie}) imply that the point of vanishing magnetization $\av{m_A} = \av{m_{B}} = 0$ develops a gradient catastrophe if
\[
t^{(c)}_{AB} = \pm \sqrt{(2 t^{(c)}_{A} - \gamma)[2 t^{(c)}_{B} -(1- \gamma)]}.
\]
When addressing the comparison of these outcomes with recent experimental results later, we will be focusing on models with no intra-party interactions (zero coupling between spins of the same type), i.e. $t_{A} = t_{B} \equiv 0$: this means that binding sites of the same type neither cooperate nor compete, while interactions between binding sites of different nature will be retained and can be both cooperative or competitive. In this case the critical point will simplify into
\[
t^{(c)}_{AB} = \pm \sqrt{\gamma (1-\gamma)}.
\]
By plugging the expression for $w$ appearing in Eq.~(\ref{biw}) into the equation~(\ref{eq:FF}), the required solution $F$ reads as follows
\begin{eqnarray}
F(x_{A}, x_{B}, t_A, t_{AB}, t_{B}) &=& \langle m_A \rangle x_{A} + \langle m_B \rangle x_{B} + \langle m_A \rangle^2 t_A + \langle m_B \rangle^2 t_B + \langle m_A \rangle \langle m_B \rangle t_{AB} \\
&-& \gamma \left[ \left (\frac{1+\langle m_A \rangle}{2} \right)\ln \left (\frac{1+\langle m_A \rangle}{2} \right) + \left (\frac{1-\langle m_A \rangle}{2} \right ) \ln \left (\frac{1-\langle m_A \rangle}{2} \right ) \right] \\
&-& (1-\gamma) \left[ \left (\frac{1+\langle m_B \rangle}{2} \right )\ln \left (\frac{1+\langle m_B \rangle}{2} \right ) + \left (\frac{1-\langle m_B \rangle}{2} \right )\ln  \left (\frac{1-\langle m_B \rangle}{2} \right ) \right].
\end{eqnarray}
Recalling \eqref{Yandm}, the relation between the average magnetizations $\langle m_A \rangle$ and $\langle m_B \rangle$ and the variables $Y_{A}$, $Y_{B}$ is given by
\be\label{eq:MandYMulti}
\langle m_A \rangle = 2 \av{Y_A} -1, \qquad \langle m_B \rangle = 2 \av{Y_B}  -1.
\ee
Hereafter, in order to lighten the notation, the bracket for $Y_{A,B}$ shall be dropped.

\begin{figure}[tbp]
\noindent \begin{centering}
\includegraphics[width=0.55\textwidth]{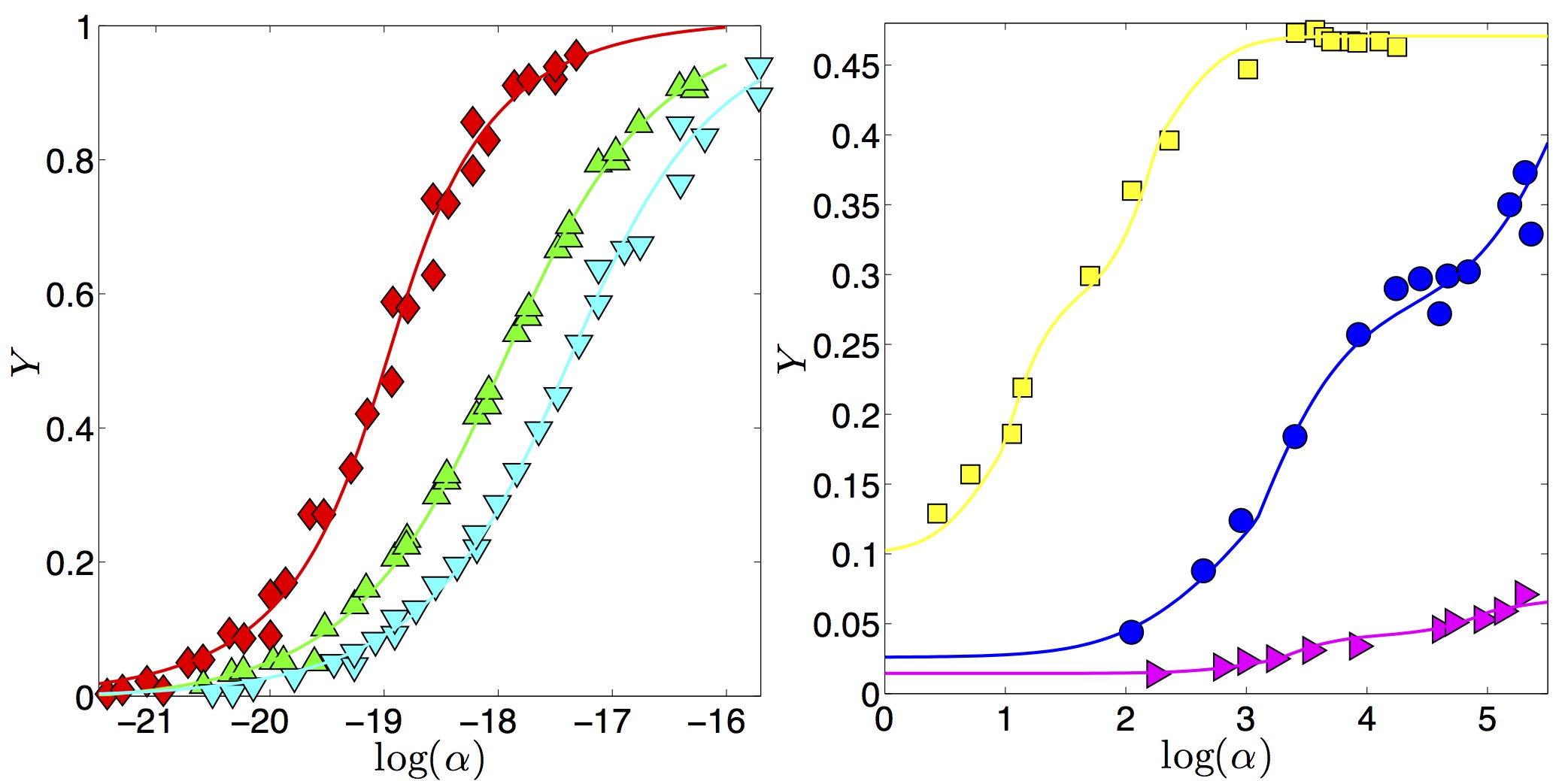}
\par\end{centering}
\caption{Examples of accordance between our theory and experimental results for positive cooperativity (left panel) and negative cooperativity (right panel). Left panel: data taken from the work by Watson et al. (see Fig.~\ref{fig:ultras} in \cite{Watson-Nature2013})
representing the fractional saturation of immobilized glucocorticoid receptor (GR) surfaces at $35 \, ^{\circ}$C as the concentration of GR DNA-binding domain (DBD) is varied; three series are considered: WT-Gilz ($\Diamond$), A477T-Pal-R ($\triangle$), and A477T-Gilz ($\bigtriangledown$). These experimental data are then fitted according to Eq.~(\ref{selfie}). The best fit coefficients are $J_{\textrm{best-fit}} \approx  0.45 \pm 0.06, 0.24 \pm 0.04, 0.27 \pm 0.04$, giving $n_H \approx 1.82 \pm 0.20, 1.32 \pm 0.07, 1.37 \pm 0.08$ (see Eq.~(\ref{hill})) to be compared with the values $n_H^{\textrm{lit}} \approx  1.83 \pm 0.28, 1.41 \pm 0.10, 1.34 \pm 0.16$ obtained in \cite{Watson-Nature2013} through a standard Hill fit.
In the right panel we consider three sets of data taken from  \cite{Garnier-Fungal1997,Glover-PNAS1975}.
Data ($\square$) taken from the work by Garnier et al. (see Fig.~\ref{fig:chemkinplots} in \cite{Garnier-Fungal1997})
represent the  fractional saturation of NAD-dependent glutamate dehydrogenase (GDH) from \emph{Laccaria
bicolor} with glutamate as substrate at pH $7.4$. These experimental data are then fitted according to Eq.~(\ref{selfie}). The best fit coefficient is $J_{\textrm{best-fit}} \approx -1.6 \pm 0.5$ giving $n_H \approx 0.38 \pm 0.09$ to be compared with the value $n_H^{\textrm{lit}} \approx  0.3$.
Data ($\circ$ and $\triangleright$) taken from the work by Glover et al. (see Fig.~\ref{fig:ruocco}a and \ref{fig:ruocco}e in \cite{Glover-PNAS1975})
represent the amino-acid uptake in \emph{B. subtilis} strain NP1 against the reciprocal of the concentration of
the transport substrate given by L-Arginine and L-Phenylanine, respectively.
These experimental data are then fitted according to Eq.~(\ref{selfie}). The best fit coefficients are $J_{\textrm{best-fit}} \approx -1.55 \pm 0.40, -2.1 \pm 0.25 $ giving $n_H \approx  0.39 \pm 0.06, 0.32 \pm 0.03$ to be compared with the value $n_H^{\textrm{lit}} \approx  0.37, 0.31$.
Notice that, in general, experimental data are reported as a function of $\alpha$ (which is the convention adopted here as well) and different systems display a different value of $\alpha_0$.
}
\label{fig:chemkinplots}
\end{figure}

\begin{figure}[htbp]
\noindent \begin{centering}
\includegraphics[width=0.7\textwidth]{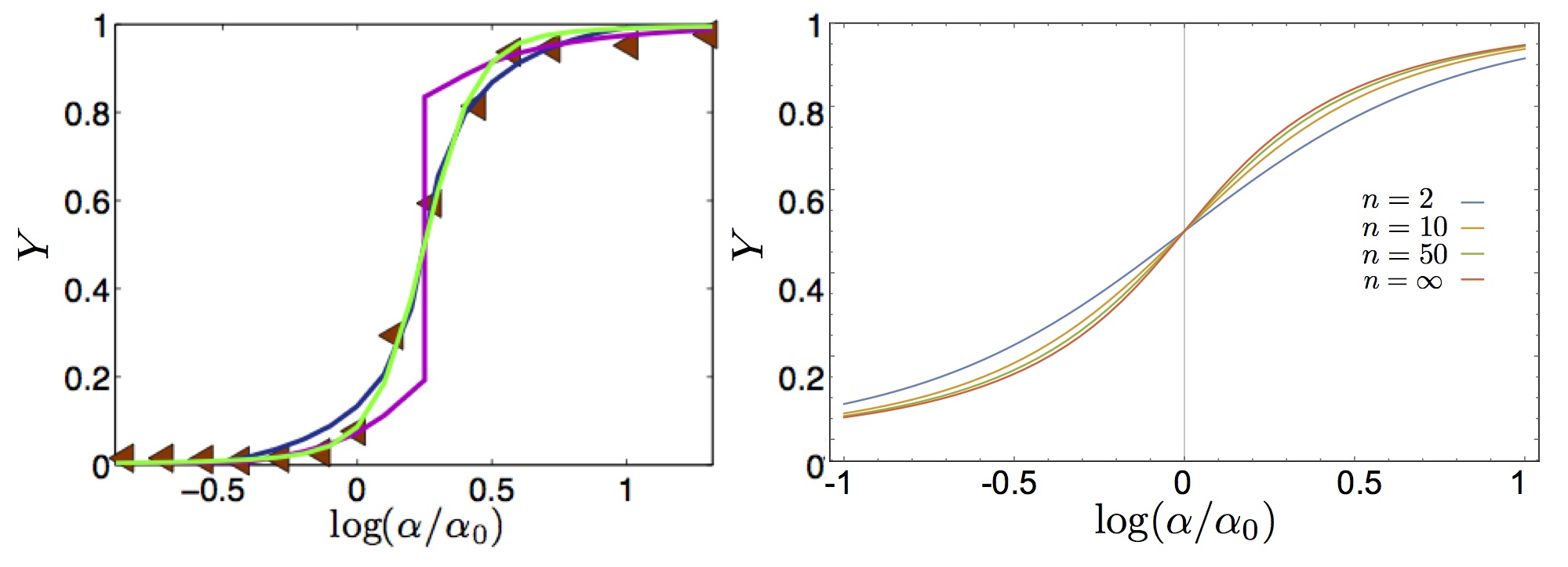}
\par\end{centering}
\caption{Left: Example of accordance between our theory and experimental results for ultra-sensitive systems. Data ($\triangleleft$) are taken from the work by Bradshaw et al. (see Fig.~3 in \cite{bradshaw}) representing CaMKII auto-phosphorylation level at equilibrium versus
[$\textrm{Ca}^{2+}$]. We report three distinct fits: the dark curve refers to Eq.~(\ref{trita}), strictly valid in the thermodynamic limit, with relatively small coupling constant (i.e., $J_{\textrm{best-fit}}<1$), hence recovering a cooperative system in the thermodynamic limit; the steep curve refers again to Eq.~(\ref{trita}) where the coupling constant is large (i.e., $J_{\textrm{best-fit}}>1$) in such a way that the response of the system gets discontinuous; the bright line refers to Eq.~(\ref{eq:ma_finite})  (where only one party is retained) valid for finite-size systems and the coupling constant is large (i.e., $J_{\textrm{best-fit}}>1$), hence recovering an ultra-sensitive system of finite size. Notice that in the latter case, given the finiteness of the system, the discontinuity is smoothened. Indeed, in the language of statistical mechanics, a genuine first order phase transition, is truly captured only in the thermodynamic limit: this mirrors an {\em infinite cooperativity} in the biochemical counterpart as it returns $n_H \to \infty$.
The relative goodness of the fits are $R^2 \approx 0.85$, $R^2 \approx 0.94$, and $R^2 \approx 0.99$, respectively, confirming an ultra-sensitive behavior. \ \ \ Right: Finite size scaling of the saturation function at different sizes (i.e $N=2$,\ 10,\ 100 and $N\to\infty$ values are shown) gives useful information on the goodness of the numerical implementation of our approach as, for small systems of concrete interest (e.g. $N<50$), the presented procedure is already feasible on standard machines, while for larger $N$ (e.g. $N> O(10^2)$) the finite-size solution already  collapses to its thermodynamic asymptotics.}
\label{fig:ultras}
\end{figure}

Before proceeding it is worth recalling that the free energy can be decomposed into $F(\beta) = - \beta \mathcal{U} + \mathcal{S}$, where the energetic and the entropic contribution are highlighted. In particular, the former
takes the form
\begin{eqnarray} \label{eq:uu}
-\beta  \mathcal{U}  &=& 2t_{AB} \left[ Y_A \cdot (1- Y_B) + Y_B \cdot (1- Y_A) \right]\\
&+& 2 Y_A \left(x_A + t_A^2\right) + 2 Y_B \left(x_B + t_B^2\right) + t_{AB} - \left( x_A + x_B + t_A^2 + t_B^2 \right).
\end{eqnarray}
Following the minimum energy principle we look for  the states that minimize $\mathcal{U}$. In particular, as long as $J \geq 0$ and $h_A h_B >0$, these states are given by $Y_A = Y_B =1$ (if $h_A, h_B >0$) and by $Y_A = Y_B =0$ (if $h_A, h_B <0$). Otherwise stated, if both the types of binding sites display a positive affinity with the ligand and reciprocal cooperativity is either absent or positive, then, according to purely energetic prescriptions, both the types of binding sites tend (not) to bind the ligand if its concentration $\alpha$ is (smaller) larger than $\alpha_0$.
\newline
Let us now consider the entropic contribution.
\be
 \mathcal{S} = - \gamma \left[ Y_A \ln Y_A + (1-Y_A) \ln (1-Y_A)\right] - (1-\gamma) \left[ Y_B \ln Y_B + (1-Y_B) \ln (1-Y_B)\right].
\ee
To see the effect of the maximum entropy principle at work instead, we differentiate $\mathcal{S}$ with respect to both $Y_A$ and $Y_B$, and impose the maximum condition, getting $Y_A = Y_B =1/2$: as expected, the states favored by the entropic term are those corresponding to half saturation for both binding sites (the most disordered states available to the system).
\newline
In general, when both entropic and energetic contributions are considered, the equilibrium states stem from an interplay between the two (as the various parameters are tuned, i.e. the noise level $\beta$, the relative size $\gamma$ of the two parties, the half-saturation reference $\alpha_0$, and the strength of the reciprocal coupling $J_A, J_B, J_{AB}$).\\
As anticipated, the maximization of $F$ allows accounting for both extremization and this leads to (see the saturation curves in Eq.s~(\eqref{selfie}) and the mapping in Eq.~(\ref{eq:MandYMulti}))
\begin{equation}\label{eq:YMultiSelfCons}
\begin{aligned}
Y_A(\alpha_A) &=& \frac12 \left \{ 1 + \tanh\left[\beta \left(J_{AB} (2 Y_B -1) + J_{AA} (2 Y_A -1)  + \frac12 \ln \left( \frac{\alpha_A}{\alpha^0_A} \right)\right) \right]\right \},\\
Y_B(\alpha_B) &=& \frac12 \left \{ 1 + \tanh\left[\beta \left(J_{AB} (2 Y_A -1) + J_{BB} (2 Y_B -1)  + \frac12 \ln \left( \frac{\alpha_B}{\alpha^0_B} \right)\right) \right]\right \}.
\end{aligned}
\end{equation}
These theoretical behaviors are successfully used to fit experimental data from positive-cooperative and negative-cooperative systems (see Fig. \ref{fig:chemkinplots}) and they are also tentatively tested on ultra-sensitive kinetics (see Fig.~\ref{fig:ultras}), although ultra-sensitivity requires particular care because of
the breakdown of the self-averaging property of the saturation function.
\noindent
In fact, within the statistical mechanical framework, we know that -away from phase transitions- the magnetization is a self-averaging order parameter, that is, when $N \to \infty$ the distribution of the magnetization becomes delta-peaked on its thermodynamic average value, namely, $\lim_{N \to \infty} P(m) \to \delta(m - \langle m \rangle)$, while when $N$ is finite this distribution is only a Gaussian (still centered on $\langle m \rangle$) whose variance vanishes as $N$ grows scaling as $1/\sqrt{N}$. However, when $J \to J_c = 1$, namely when the system exhibits a phase-transition (or when a shock develops in the language of non-linear waves) the variance diverges as $N \to \infty$. In practice, this means that, away from phase transitions, $\lim_{N \to \infty} (\langle m^2 \rangle - \langle m \rangle^2) = 0$ but, exactly on the critical point (i.e. $J=J_c$ and $h=0$), this relation does not hold any longer and fluctuations in the order parameter grows indefinitely with the size.
\newline
Analogous arguments apply to $Y$ as well, since $Y$ and $m$ are linearly related. Then, the criticality (whose signatures emerge even in real systems as $N$ gets large, $\alpha = \alpha_0$ and $J \to J_c$) has important consequences also from a practical perspective. For any $J> J_c$, as larger and larger systems are considered, the saturation function $Y(\alpha)$ displays a discontinuity at $\alpha = \alpha_0$: at that point the system jumps from a (almost) empty state to a (almost) fully-occupied state (the jump is meant as a real discontinuity of $Y(\alpha)$ at $\alpha=\alpha_0$ only for $N \to \infty$ and the degree of occupation of the two extremal states depends on the level of noise). The lack of smoothness in the function $Y(\alpha)$ makes the application of regression techniques awkward. Thus, despite from a biochemical perspective ultra-sensitive reactions are {\em only} particularly strong cooperative reactions, from a mathematical perspective these two types of reactions are actually very different.
\newline
In the next subsection we will develop the small $N$ theory that offers a more refined level of description for these critical systems, and, in particular, we will succeed in evaluating the ``jump'' that $Y(\alpha)$ experiences at $\alpha=\alpha_0$ at finite volume $N$ using shock theory: this will result in a practical instrument that can be used in modern experiments on small systems reaction kinetics.

\emph{\textbf{Finite-size solution.}}
Here we assume that $N$ is finite and fixed and we study the exact solution of the system~(\ref{Zpdes}) with the initial condition~(\ref{Zpdeinit}).
Let us remark that the initial condition~(\ref{Zpdeinit}) can be equivalently written as
\begin{equation}
Z_{0} = \sum_{k=0}^{N_{A}} \;  \sum_{l=0}^{N_{B}} \binom{N_{A}}{k} \binom{N_{B}}{l} \; e^{-\frac{N}{N_{A}} (N_{A} - 2 k) x_{A}} \; e^{-\frac{N}{N_{B}} (N_{B} - 2 l) x_{B}},
\end{equation}
where, we recall, $N =N_{A} + N_{B}$.
Observing that the initial condition is separable in the variables $x_{A}$ and $x_{B}$, we then look for separable solutions of the form
\begin{equation}
\label{Zsep}
Z(x_{A},x_{B},t_{A},t_{AB},t_{B}) = \sum_{k=0}^{N_{A}} \;  \sum_{l=0}^{N_{B}} \binom{N_{A}}{k} \binom{N_{B}}{l} \; a_{k,l}(t_{A}) b_{k,l}(t_{AB}) c_{k,l}(t_{B}) e^{-\frac{N}{N_{A}} (N_{A} - 2 k) x_{A}} \; e^{-\frac{N}{N_{B}} (N_{B} - 2 l) x_{B}}.
\end{equation}
Substituting the previous expression into the equations~(\ref{Zpdes}) we obtain
\begin{align*}
a_{k,l} = e^{\frac{N}{N_{A}^{2}} (N_{A} -  2 k)^{2} t_{A} }, \qquad  b_{k,l} = e^{\frac{N}{N_{A} N_{B}} (N_{A} -  2 k) (N_{B} -  2 l) t_{AB} }, \qquad c_{k,l} = e^{\frac{N}{N_{B}^{2}} (N_{B} -  2 l)^{2} t_{B} }.\qquad
\end{align*}
The required solution is
\begin{align*}
Z(x_{A},x_{B},t_{A},t_{B},t_{AB}) =& \sum_{k=0}^{N_{A}} \sum_{l=0}^{N_{B}} \binom{N_{A}}{k} \binom{N_{B}}{l} \;  \exp \left[ \Lambda_{k,l}(x_{A},x_{B},t_{A},t_{B},t_{AB})\right],
\end{align*}
where
\begin{align*}
\Lambda_{k,l}(x_{A},x_{B},t_{A},t_{B},t_{AB}) &=\frac{N}{N_{A}^{2}} (2k - N_{A})^{2} t_{A}+ \frac{N}{N_{A} N_{B}} (2k - N_{A}) (2 l - N_{B}) t_{AB}   + \frac{N}{N_{B}^{2}} (2l - N_{B})^{2} t_{B}  \\
&+ \frac{N}{N_{A}} (2k - N_{A}) x_{A}  + \frac{N}{N_{B}} (2l - N_{B}) x_{B}.
\end{align*}
In particular, the finite-size magnetizations
\[
\langle m_{A} \rangle = \frac{1}{N} \der{\log Z}{x_{A}} \qquad \langle m_{B} \rangle = \frac{1}{N} \der{\log Z}{x_{A}}
\]
take the following explicit expressions
\begin{gather}
\label{eq:ma_finite}
\begin{aligned}
&\langle m_{A} \rangle =  \frac{1}{N_{A}} \frac{1}{Z} \left \{ \sum_{k=0}^{N_{A}} \sum_{l=0}^{N_{B}} \binom{N_{A}}{k} \binom{N_{B}}{l} (2k - N_{A} ) \exp \left[ \Lambda_{k,l}(x_{A},x_{B},t_{A},t_{B},t_{AB})\right] \right \}, \\
&\langle m_{B} \rangle =  \frac{1}{N_{B}} \frac{1}{Z} \left \{ \sum_{k=0}^{N_{A}} \sum_{l=0}^{N_{B}} \binom{N_{A}}{k} \binom{N_{B}}{l} (2l - N_{B}  ) \exp \left[ \Lambda_{k,l}(x_{A},x_{B},t_{A},t_{B},t_{AB})\right] \right \}.
\end{aligned}
\end{gather}
Notice that, by fixing $t_{A} = t$, $t_{AB} = t_{B} = 0$, $x_{B}=0$ and $N_{A} = N$, we recover the finite-size solution to the Curie--Weiss model, as expected. Also, it is immediate to recognize that $\exp [\Lambda_{k,l}(x_{A},x_{B},t_{A},t_{B},t_{AB})] / Z$ plays the role of the finite-size canonical probability for the configuration with $k$ and $l$ bounded sites out of $N_A$ and of $N_B$, respectively (corresponding to a magnetizations $2k - N_A$ and $2l - N_B $).
\newline
Fitting Eq.s~(\ref{eq:ma_finite}) on ultra-sensitive kinetics (as shown in Fig.~\ref{fig:ultras}) produces sensibly better results than using the infinite limit counterpart coded by Eq.s~\ref{eq:YMultiSelfCons}.

%

\section*{Implications of the theory for small systems}

The successful comparison between the experimental saturation plots and the predictions from self-consistencies obtained in the previous section provides a sound check of the analogy developed. Now, we aim to go further to recover novel emerging phenomena, whose experimental evidence is well documented.
In particular, we now focus on cooperative-like and bistable behaviors induced by noise and stochastic effects \cite{Warren,Kauffman-JIntBiol2006,Lipshtat-PRL2006,kardar-bistability,Leibler-Nature2000,transcriptional-scenario}. Next, this phenomenology is further investigated in its relationship with quorum sensing \cite{Kardar-PNAS2013} and  we will finally offer a cybernetic interpretation of this kind of process: this  will allow  us to provide a natural and robust explanation about how, during a biochemical reaction, the {\em signal} (carried by the input) is amplified, while an eventual {\em noise} becomes spontaneously attenuated.

\emph{\textbf{Noise-induced cooperativity and bistability.}}
It is well-known that cooperative binding in real systems (i.e. at finite sizes) induces a bistable behavior, which we call \emph{cooperativity-induced bistability}.
On the other hand, growing attention has been recently captured by the evidence that \emph{bistability} can also occur  \emph{without} cooperativity  (i.e. it may happen in systems whose binding sites do not interact at all, simply as a consequence of the intrinsic stochasticity, see e.g. \cite{Lipshtat-PRL2006} and references therein). We call this phenomenon \emph{noise-induced bistability}.
Indeed, these two kinds of bistability display a significant empiric resemblance that makes one speculate that stochasticity can play a a role in the effective cooperativity. However, the two phenomena are conceptually different, as it can be inferred from the present statistical mechanical treatment.
The simplest way to do this is by starting from the mono-partite system discussed in Sec.~$2$ and compare  the self-consistent expression \eqref{trita} with the saturation function of cooperative systems \eqref{collina}. Both expressions are recalled hereafter for clarity:
\ba \label{eq:SCagain}
Y (\alpha) &=&  \frac{1}{2} \left \{ 1 + \tanh \left[ \beta J (2Y-1) + \frac{\beta}{2} \log \left( \frac{\alpha}{\alpha_0} \right) \right] \right \}.\\
\label{eq:collina_again}
Y (\alpha) &=& \frac{K \alpha^{n_H}}{1+\alpha^{n_H}},
\ea
Notice that in Eq.~(\ref{eq:SCagain}) the explicit $\beta$ is retained (i.e., the rescaling $\beta J \to J$ and $\beta h \to h$ is no longer applied) since here we are focusing on the role of $\beta$.\\
Now, it is easy to see that stochastic effects may drift the system away from Michaelis--Menten behavior (toward a cooperative-like one), even if there is no cooperation among binding sites. In fact, by assuming that the coupling is strictly zero, i.e. by setting $J \equiv 0$ in the self-consistency (\ref{eq:SCagain}), we get
\be
Y = \frac{1}{2} \left \{ 1 + \tanh \left[  \frac{\beta}{2} \log \left( \frac{\alpha}{\alpha_0} \right) \right] \right \}
\ee
whence, recalling that $\tanh r=1 -2/(e^{2r}+1)$, we have
\be
Y(\alpha) = \frac{\left( \frac{\alpha}{\alpha_0} \right)^{\beta}}{1 + \left( \frac{\alpha}{\alpha_0} \right)^{\beta}}.
\ee

The last expression coincides with the Michaelis-Menten law $\eqref{MM}$ as long as $\beta=1$ and $\alpha_0$ coincides with the Michaelis constant (defined as the ratio between dissociation and association constants related to the reaction considered, which corresponds to
the concentration of the ligand at which the reaction rate is half its maximum value). More generally, letting $\beta$ vary (hence rising or lowering stochastic effects in the system), we can reshuffle the previous expression as
\be \label{MMreloaded}
Y(\alpha,\beta) = \frac{\alpha^{\beta}}{\alpha_0^{\beta} + \alpha^{\beta}},
\ee
which recovers the Hill law \eqref{collina} as long as we take $K^{-1} = \alpha_0^{\beta}$. Focusing on the dependence of $Y$ on the concentration $\alpha$, we find that $\beta$ plays the same role as the Hill coefficient: the smaller the stochasticity (i.e. the larger $\beta$) the more ``cooperative'' the system.
The reason for this behavior is apparent in the statistical mechanical picture, as low levels of noise make the spins of the system align faster (driven by the external field, rather than by themselves) by reducing overall the fluctuations. Of course, the displayed behavior depends \emph{globally}, rather than locally, on the system energy, i.e. it may not be ascribed to the mutual interaction of sites, as it is the case for truly cooperative systems.

\begin{figure}[htbp]
\noindent \begin{centering}
\includegraphics[width=0.5\textwidth]{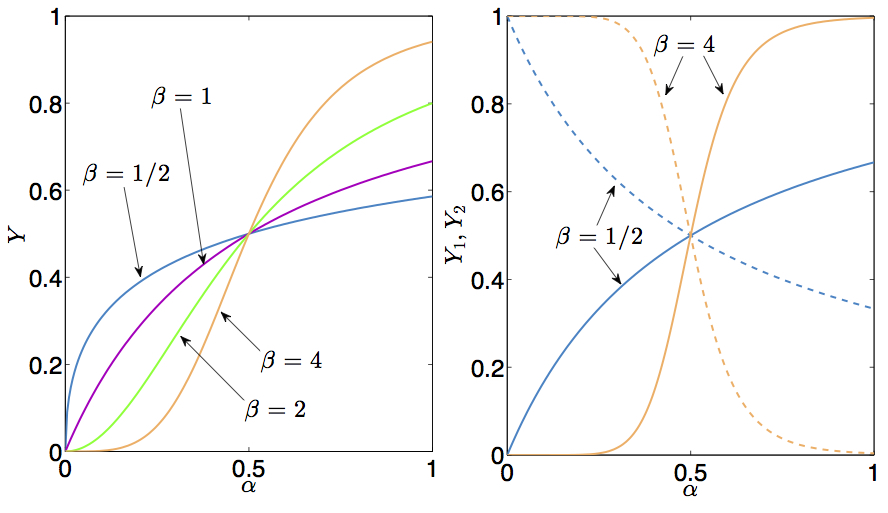}
\par\end{centering}
\caption{(Left Panel) Noise induced cooperativity. The saturation function of a single-type site reaction $Y(\alpha)$ is shown for various values of the noise $\beta$, according to Eq.~(\ref{MMreloaded}). At high level of noise the Michaelis--Menten envelop is preserved (the curve at $\beta=1/2$), while for higher values of $\beta$ -- hence for smaller noises -- a sigmoidal shape slowly emerges (i.e. for $\beta=2,\ 4$). (Right Panel) Noise induced bistability. The saturation function of a two-type site reaction $Y_\A(\alpha),\ Y_\B(\alpha)$ is shown for various values of the noise~$\beta$, according to Eq.~(\ref{eq:LDS}). Again, while for small values (i.e. $\beta = 1/2$) the system behaves as the superposition of two independent Michaelis--Menten reactions, for higher values of $\beta$ (i.e. $\beta=4$) bistability clearly appears in the system. Here we used $\alpha_A = \alpha_B$ and $\alpha_0^A = \alpha_0^B= 1/2$.}
\label{fig:NIC}
\end{figure}

In Fig.~\ref{fig:NIC} (left panel) we show the qualitative behavior of the saturation function $Y$ as a function of the concentration $\alpha$ and with noise parameter $\beta$, according to Eq.~(\ref{MMreloaded}).

Let us now proceed by showing that noise can even induce bistability. To this aim, we need to extend this approach to systems built of by two-parties (namely to the general model coded in Eq.~(\ref{bizione})); despite being mathematically more involved, the reasoning for multipartite systems goes as much as the same as above. Indeed, profiting the mappings \eqref{eq:MandYMulti}, the partial saturation curves may be shown to represent Hill law for each party also in the case of two (or several) interacting sites.

To see this, let us set
\begin{align*}
Y_{\A}(\alpha) \equiv & \frac{ \left \langle \sum_{i=1}^{N_\A} (\sigma_i^\A +1)  \right \rangle }{2 N_\A} & Y_\B(\alpha)\equiv& \frac{ \left \langle \sum_{i=1}^{N_\B} (\sigma_i^\B +1)  \right \rangle }{2 N_\B} ,
\end{align*}
where $\alpha_\X$ denotes the concentration of free molecules for the party $\X=\A,\B$ and $\alpha_0^\X$ denotes the reference value representing equal probability of being occupied and unoccupied for the party $\X$.

Assuming again $J=0$ and $N_\A=N_\B=N/2$ for the sake of simplicity, the event of each site being bound (respectively unbound) is independent of the state of the other sites, thus, denoting with $\sigma^\X_i$ the $i^{\textrm{th}}$ spin ($i=1,\dotsc,N_\X$) of the party $\X$, the total concentration 
\begin{align*}
\alpha_{\textrm{tot}}\propto \frac{P\tonde{\sigma_{i_1}^\A=+1, \sigma_{i_2}^\B=+1}}{ P\tonde{\sigma_{i_1}^\A=-1,\sigma_{i_2}^\B=-1}}
\end{align*}
factorizes due to the independence of the probabilities (a major advantage of the mean-field approach) as
\begin{align*}
\alpha_{\textrm{tot}}\propto \frac{P\tonde{\sigma_{i_1}^\A=+1} \cdot  P\tonde{\sigma_{i_2}^\B=+1}}{ P\tonde{\sigma_{i_1}^\A=-1} \cdot  P\tonde{\sigma_{i_2}^\B=-1}} ,
\end{align*}
whence $\alpha_{\textrm{tot}}=\alpha_\A\cdot\alpha_\B$ (and the same conclusion may be drawn for the total reference value $\alpha_0^{\textrm{tot}}$). As a consequence of the superposition principle for the external fields and of \eqref{eq:ha}, we then have
\begin{align*}
h_\A+h_\B=h_{\textrm{tot}}=\frac{1}{2}\log\tonde{\frac{\alpha_{\textrm{tot}}}{\alpha_0^{\textrm{tot}}}}=\frac{1}{2}\log\tonde{\frac{\alpha_\A}{\alpha_0^\A}\cdot\frac{\alpha_\B}{\alpha_0^\B}}=\frac{1}{2}\log\tonde{\frac{\alpha_1}{\alpha_0^\A}}+\frac{1}{2}\log\tonde{\frac{\alpha_2}{\alpha_0^\B}}
\end{align*}
whence, having set $N_\A=N_\B$, we can put by symmetry (see Eq.~\eqref{eq:ha})
\begin{align*}
h_\A\equiv& \frac{1}{2}\ln\tonde{\frac{\alpha_\A}{\alpha_0^\A}} & h_\B\equiv& \frac{1}{2}\ln\tonde{\frac{\alpha_\B}{\alpha_0^\B}}.
\end{align*}

Exploiting again the assumption $J=0$, that is, in the absence of cooperativity Eq. \eqref{eq:YMultiSelfCons}  reads as
\begin{align*}
Y_\A(\alpha_\A)=&\frac{1}{2}\set{1+\tanh\quadre{\beta\ln\tonde{\frac{\alpha_\A}{\alpha_0^\A}}}} &
Y_\B(\alpha_\B)=&\frac{1}{2}\set{1+\tanh\quadre{\beta\ln\tonde{\frac{\alpha_\B}{\alpha_0^\B}}}}
\end{align*}
whence, replacing the hyperbolic tangent with its exponential expression, we recover the partial Hill laws
\be
 \label{eq:LDS}
Y_\A(\alpha_\A,\beta)=\frac{\alpha_\A^{2\beta}}{\tonde{\alpha_0^\A}^{2\beta}+\alpha_\A^{2\beta}},  ~~~~ ~~~~
Y_\B(\alpha_\B,\beta)=\frac{\alpha_\B^{2\beta}}{\tonde{\alpha_0^\B}^{2\beta}+\alpha_\B^{2\beta}}
\ee
for the single parties. Notice that, in order for the analogy with eq.~\eqref{collina} to be preserved, the partial Hill coefficients coincides now with $2\beta$, rather than $\beta$, for each single party consists of precisely one half of the total number of spins. This is indeed perfectly consistent with our previous remark on the global dependence on the noise $\beta$ in the case of noise-induced ``cooperativity'' and is indeed in sharp contrast with the property of the truly cooperative behavior to be invariant with respect to the system size.

In order to graphically illustrate the properties of the partial saturation curves $Y_\X$ with respect to the noise $\beta$, fix e.~g. $h_\A=h_\B=h_{\textrm{tot}}/2$, so that $\alpha_\B= \alpha_\A \alpha_0^\B / \alpha_0^\A $ and $\alpha_{\textrm{tot}}= (\alpha_\A)^2 \alpha_0^\B / \alpha_0^\A $. By fixing a value of $\alpha_{\textrm{tot}}$ (or, which is the same, the values of $\alpha_0^\X$), we can interpret the $\alpha_\X$'s to be relative concentrations with respect to $\alpha_{\textrm{tot}}$, hence rewriting $Y_\B(\alpha_\B)=Y_\B(\alpha_{\textrm{tot}}/\alpha_\A)\equiv Y_\B(\alpha_\A)$ for $\alpha_\A\in(0,1)$. This stochastic-driven behavior for two-party systems is shown in  Fig.~\ref{fig:NIC} (right panel).
\begin{figure}[htbp]
\noindent \begin{centering}
\includegraphics[width=0.6\textwidth]{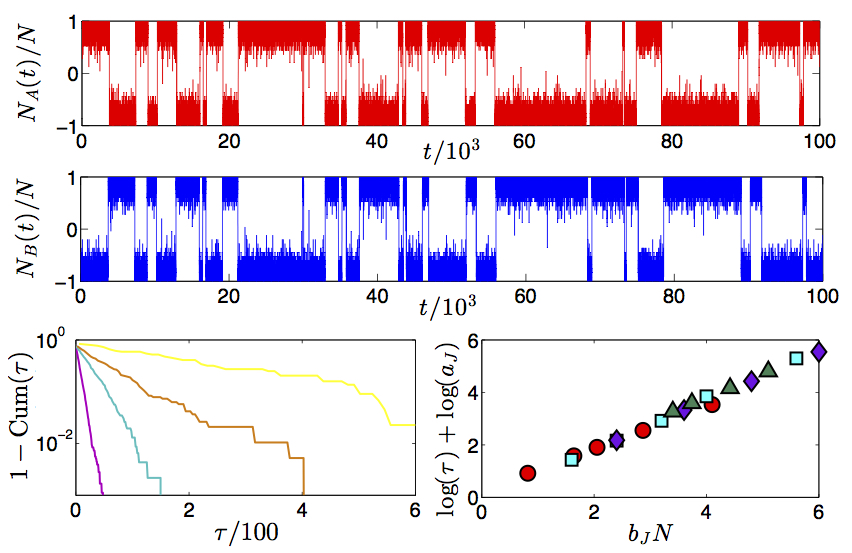}
\par\end{centering}
\caption{Upper panel:  example of historical series for a toggle switch we simulated through a bipartite spin system with overall $N=40$, $N_A=N_B=20$ and $J_{AB}=-2.8$, $J_A=J_B=0$. The behavior of this  {\em stochastic flip-flop} is as much the same as that of classical toggle switches, as it is evident comparing this plot with Fig. 3 of \cite{Lipshtat-PRL2006} or Fig 3(a) of \cite{Warren3}.
Lower panels: the switching time for various couplings' strength is analyzed. The complementary of the cumulative distribution $\textrm{Cum}(\tau)$ is shown in the left for $J_{AB}=-3.0, -2.8, -2.7, -2.5$ (ordered from the least to the most steep curve). The scaling law $\tau = a_J \exp(b_J N)$ is checked in the right panel, where, again different coupling strengths are considered [$J_{AB}=-3.0 (\diamond) , -2.8 (\triangle), -2.7 (\square), -2.5 (\circ)$];  to be compared to  \cite{Warren3} Fig.~3b.}
\label{fig:serie_temporali}
\end{figure}

\begin{figure}[tbp]
\noindent \begin{centering}
\includegraphics[width=0.7\textwidth]{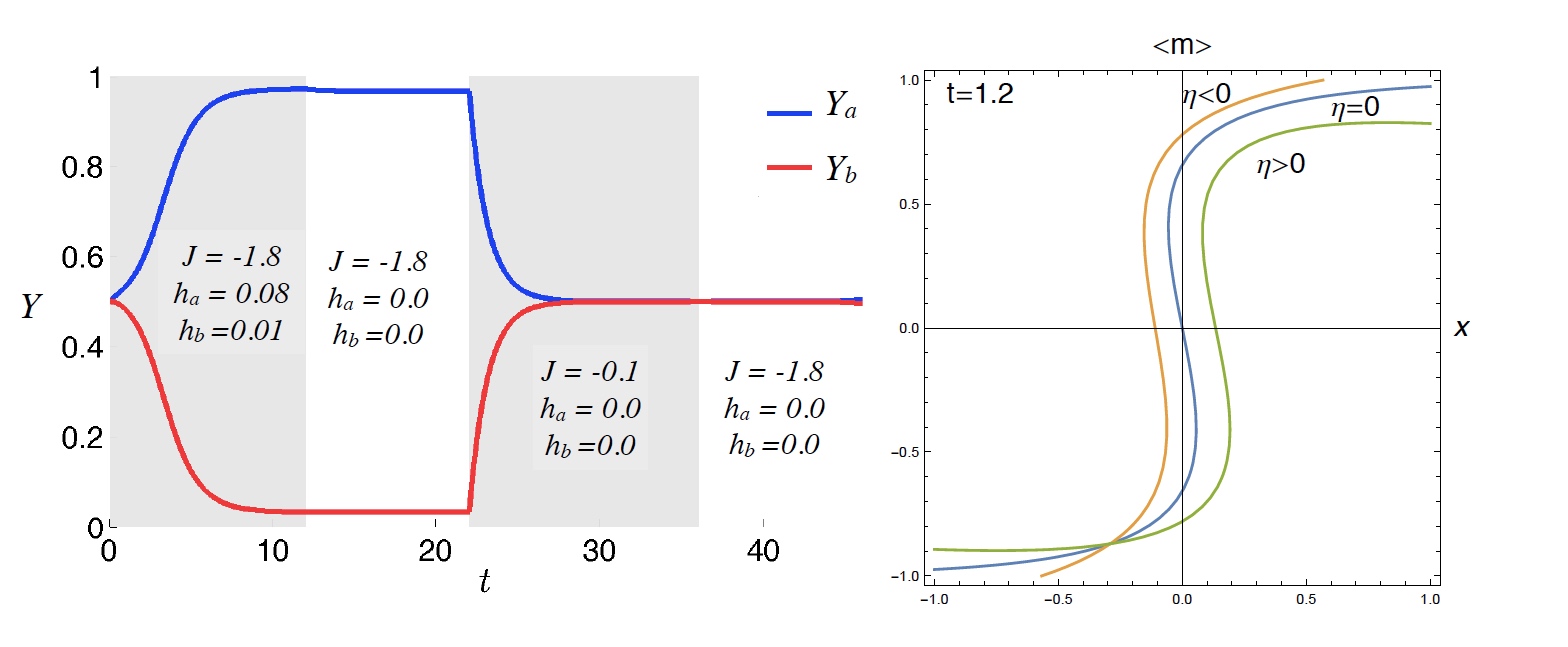}
\par\end{centering}
\caption{Left: Temporal behavior of the outputs $Y_\A$ and $Y_\B$ for the two parties. Along the time, the parameters $J$, $h_\A$ and $h_\B$ are varied as specified in the figure. In this way, four distinct regimes can be outlined: at the beginning both parties display half saturation and negative cooperation, being subject to small, positive fields and, as a result, the output associated to the party experiencing the larger field grows while the other decreases; then, both fields are set to zero and the outputs remains close to $1$ and to $0$, respectively; next, the coupling between the two parties is reduced and the related output collapse to $1/2$; now, even if the coupling is restored to large, negative values the two outputs remain null. This phenomenology recovers the original picture by Gardner et al. (see \cite{Gardner-Nature2000}, Fig.~\ref{fig:ultras}). Right: First order corrections to the infinite volume expression for the transfer function of the Curie--Weiss model: $\eta$ encodes for random thermal fluctuations that, in this context, are not washed away and actually break locally (in time) the symmetry between the two free energy minima allowing the system to oscillate between them (the stronger the thermal noise the faster the hopping rate between the two minima).}
\label{Gardner}
\end{figure}

The physical reason for this switches lies in the intrinsic ergodicity of any dynamics involving systems with finite size:  considering the simplest bistable system, its two free energy minima are separated by a maximum (i.e., a {\em barrier}) whose height is $N \delta F$ and the characteristic time $\tau$ for the system to move from one minimum to another typically scales as $\tau \propto \exp( \delta F)$ thus, solely in the thermodynamic limit, this barrier becomes infinite and the system gets {\em trapped} forever into one of these minima (and ergodicity becomes broken): this is clearly discussed from a  biochemical perspective in e.g., \cite{Lipshtat-PRL2006,Warren3} (and references therein). We checked that our theory correctly reproduces these spontaneous switches in time, as reported in Fig.s~\ref{fig:serie_temporali} and~\ref{Gardner} (left panel) where we recover the original plots shown  \cite{Lipshtat-PRL2006} using the same parameters.

In our vocabulary, this happens because the asymptotic evaluation of the magnetization already for the simplest bistable mono-partite spin model (i.e., a Curie--Weiss model) leads to the expansion of the form
\begin{equation}
\av{m} = \xi + O\left(\frac{1}{\sqrt{N}} \right)
\end{equation}
for large $N$, where $\xi$ is the solution to the self-consistency equation
\begin{equation}
\xi = \tanh(t \xi  + x)
\label{selfcon}
\end{equation}
such that the free energy attains its minimum. Away from the thermodynamic limit, keeping only the first order correction to finite volume, we can approximate
\begin{equation}
\av{m} \simeq \xi + \frac{\eta}{\sqrt{N}}
\label{mexp}
\end{equation}
and $\eta$ can be interpreted as a stochastic contribution. Indeed, the magnetization $\av{m}$ has zero-variance in the limit $N \to \infty$ only (or in the pathological case of zero noise $\eta \equiv 0$). Solving the equation~(\ref{mexp}) for $\xi$ and substituting in~(\ref{selfcon}), we obtain (up to $O(1/N)$)
\[
\av{m} = \tanh(t \av{m} + x) + \left [(t+1) - t m^{2} \right]  \frac{\eta}{\sqrt{N}}.
\]
Thus, as shown in Fig.~\ref{Gardner} (right panel) a non-vanishing $\eta$ breaks the symmetry of the isothermal curve for the average magnetization $\av{m}$. In absence of external field, i.e. $x=0$, a positive (negative) value of $\eta$ selects a negative (positive) solution $\av{m}$, hence random fluctuations of $\eta$ are responsible for possible switches of the magnetization thus effective stochastic bistabilities typical of toggle switches.

\begin{figure}[h]
\noindent \begin{centering}
\includegraphics[width=0.7\textwidth]{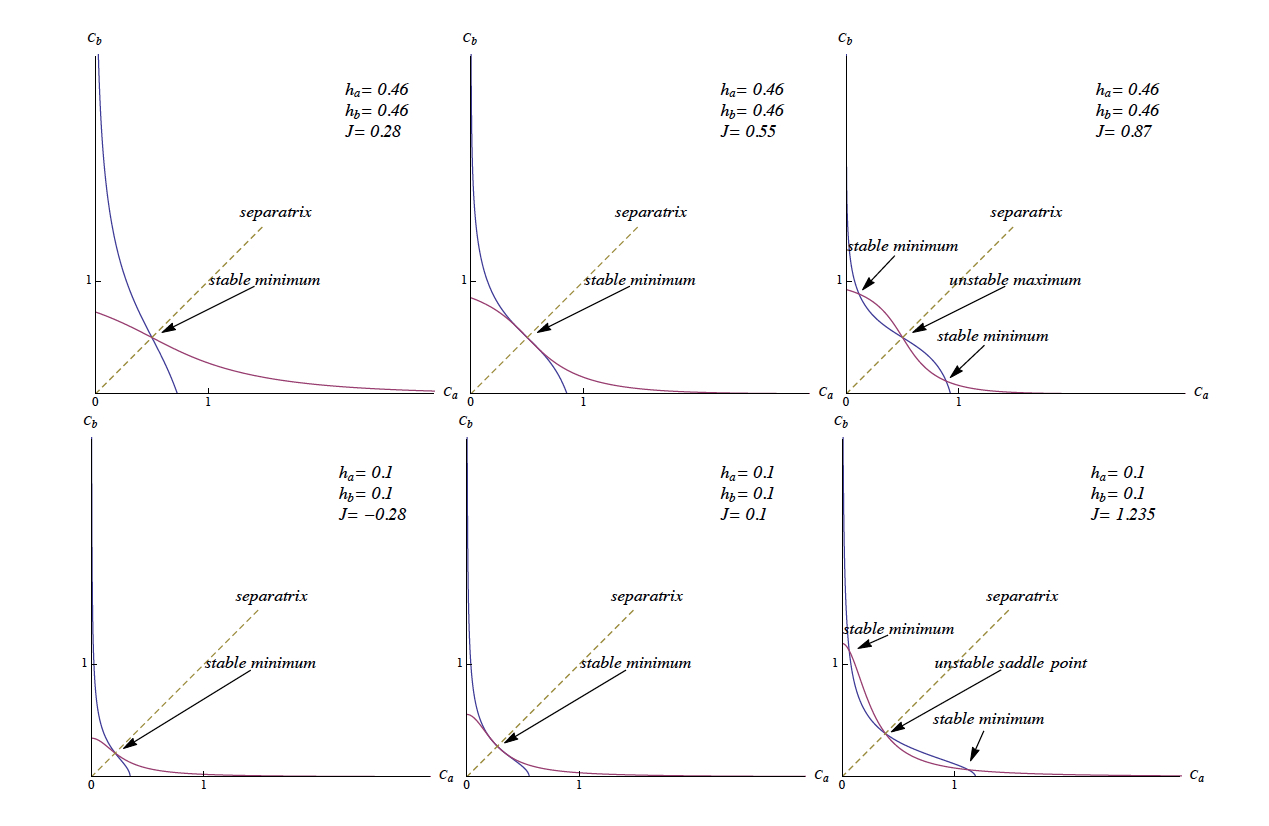}
\par\end{centering}
\caption{Upper panel: Null-clines of Langevin equations system \eqref{Lange0}, displaying equilibria and their behavior, with varying $J$.
Lower panel: Null-clines of the system \eqref{eq:typicalToggle}, roughly displaying the same behavior as \eqref{Lange0} (to be compared to \cite{Leibler-Nature2000}, Fig. 2). Notice that in the present case, the concentrations $C_\A, C_\B$ are not normalized as they previously were, so that, after a suitable renormalization, the equilibria belong in fact to region $[0,1]\times [0,1]$. 
}
\label{fig:nullclines}
\end{figure}

To further check that our theory correctly reproduces such a flip-flop (i.e. switch) like behavior, we can study the null-clines of  the simplest Langevin dynamics coupled to our system (\ref{bizione}), namely
\begin{eqnarray}\label{Lange0}
\tau \frac{d \langle m_\A \rangle}{dt} &=& - \frac{\partial H(m_\A,m_\B \mid J_{\A\B}, h_\A,h_\B)}{\partial m_\A} + \eta_\A = - \av{m_A} + \tanh\left(-\beta J \av{m_B} + h_A \av{m_A}\right) + \eta_A,\\ \label{Lange1}
\tau \frac{d \langle m_\B \rangle}{dt} &=& - \frac{\partial H(m_\A,m_\B \mid J_{\A\B}, h_\A,h_\B)}{\partial m_\B} + \eta_\B  = - \av{m_B} + \tanh\left(-\beta J \av{m_A} + h_B \av{m_B}\right)+ \eta_B.
\end{eqnarray}
where $\tau$ is the typical time constant (see Fig.~\ref{fig:serie_temporali}), and the randomness tuned by $\eta$ is standard white noise, namely $\langle \eta_\A(t) \eta_\B(t')\rangle \propto \delta(\A-\B)\delta(t-t')$: keeping $C_\A$, and $C_\B$ to label the two reactant's concentrations as in the original paper \cite{Leibler-Nature2000}, we can compare the null-clines of our system to those pertaining to the following archetypal toggle
\begin{align}\label{eq:typicalToggle}
\dot C_\A=\frac{g_\A}{1+ C_\B^{n_H^\B}}-d_\A C_\A\\
\dot C_\B=\frac{g_\B}{1+ C_\A^{n_H^\A}}-d_\B C_\B,
\end{align}
where we have kept $C_\A$ and $C_\B$ to label the un-normalized reactant's concentrations. The corresponding by now well-known results are illustrated in Fig.~\ref{fig:nullclines} (lower panel) for the case $n_H^\A=n_H^\B=2$.
\newline
Letting $F_{\A}(m_\A,m_\B)\equiv -m_\A+\tanh(-Jm_\B+h_\A m_\A)$ (and $F_\B$ analogously), the system's (in-)stability at equilibria may be checked with the sign of the differential's eigenvalues
\begin{equation*}
D\equiv \begin{pmatrix}
\partial_{m_\A} F_\A & \partial_{m_\B} F_\A \\
\partial_{m_\A} F_\B & \partial_{m_\B} F_\B
\end{pmatrix},
\end{equation*}
an equilibrium point being a minimum (i.e. stable) if both the eigenvalues are positive and a maximum (i.e. unstable) if both of them are negative, or otherwise a saddle point (unstable) if they are opposite in sign.

\emph{\textbf{Condorcet theory: stochastic signal amplification and noise suppression.}}
The Condorcet theorem was originally stated within a political context and concerns the relative probability of a given group of individuals arriving at a correct (binary, i.e. YES-NO) decision.
The assumptions underlying the simplest version of the theorem is that a group wishes to reach a decision by majority vote. One of the two outcomes of the vote is correct, and each voter has an independent probability $p$ of voting for the correct decision. The theorem states that if $p>1/2$ (i.e., each voter is more likely to vote correctly), then adding more voters to the pool increases the probability that the majority decision is correct. In the thermodynamic limit, the probability (that the majority votes correctly) approaches $1$ as the number of voters diverges.
On the other hand, if $p< 1/2$ (i.e., each voter is more likely not to vote correctly), then adding more voters makes things worse.

Condorcet theorem is therefore a majority rule that can be used as a tool to infer collective reliable predictions about the better of two options and, in recent times, deep analogies between Condorcet theory in social systems and {\em quorum sensing} in biological information processing systems have been developed: for instance, many species of bacteria use quorum sensing to coordinate gene expression according to the density of their local population \cite{quorum-bacteria}, and decisions within the adaptive response of the immune system in mammals requires the implementation of quorum sensing by lymphocytes \cite{Kardar-PNAS2013,Agliari-JTB2015} (that globally {\em vote} if attacking or resting when a novel antigen is presented to them and dialogue among themselves respecting the rules of biochemical reactions). In any case, this distributed decisional capability should emerge as a natural consequence of the underlying reaction kinetics of all the involved agents, thus we should be able to understand its genesis from our general treatment.

\begin{figure}[htbp]
\noindent \begin{centering}
\includegraphics[width=0.7\textwidth]{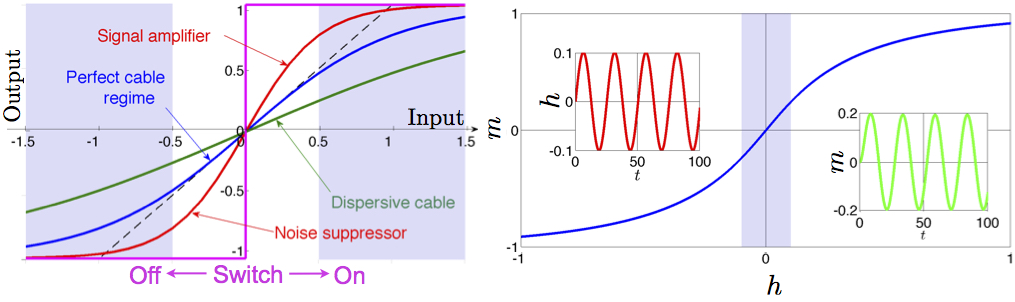}
\par\end{centering}
\caption{Left panel: Transfer function for cooperative systems (i.e. chemical cooperative kinetics, physical ferromagnets, electronic operational amplifiers). Curves pertaining to different interaction strength among the system's units (namely pertaining to a different Hill coefficient, or a different ferromagnetic coupling, or a different gain, according to the context considered) are shown in different colors. The dashed line highlights the linear behavior expected in the region of small signal, namely for a perfect cable. As the interaction strength gets larger we eventually get a switch. The Condorcet scenario results in both signal amplification at high ligand concentration as well as noise suppression at low ligand concentration. Right panel: Amplification scheme. In the main plot the linear regime is highlighted. When the input (i.e., the external field $h$, left inset) is varied within this region the resulting output (i.e., the magnetization $m$, right inset) varies according to a linear relation. Outside the linear regime the input-output relation can exhibit distortions.}
\label{fig:kardar}
\end{figure}

Note at first that this mechanism is also consistent with the biologically-adapted \cite{zhang1,ABBDU-SciRep2014} electronic picture of {\em signal amplification} and {\em noise suppression} (where, in this analogy, the meaning of the signal is the correct vote and that of noise is the wrong one) as, actually, this is not too far from the Shannon coding theorem in Cybernetics: roughly speaking the latter states that, even if there is a huge noise on a cable (i.e. in the {\em system}), it is enough that the probability to send successfully a message through it remains strictly greater than one-half to be sure that the the whole information that crossed the system cannot get lost (with large enough trial samples $N \to \infty$). Indeed in all these saturable systems (as also voter-like models a' la Condorcet are saturable systems by definition if the vote is binary) when collective features are at work, the correct output ``yes'' emerges neatly even at a relatively mild input, and the wrong output ``no'' is cleverly avoided even in the presence of significant noise, as shown in Fig.~\ref{fig:kardar}.

This kind of phenomenology is naturally captured within our chemical kinetics framework: in fact, in our system the stimulus (i.e., the logarithm of the ligand concentration) is provided by the field $h$ and, according to its effective magnitude, it is expected to return a bound (``yes'') or a not-bound (``no'') state for the ligands such that, if the stimulus is poor (i.e., chemical noise), it is suppressed in the output, while if the stimulus is relatively consistent (i.e., chemical signal), it gets amplified in the output, as discussed in detail in Fig.~\ref{fig:kardar}. Clearly the larger the coupling value $J>0$ (or, alternatively, the higher the Hill coefficient of the reaction), the stronger the resulting amplification (see also Eq.s (\ref{compare})). This point is further deepened hereafter. If we look at the expression for the system's energy (Eq.~(\ref{eq:uu})), a key point is that when $J>0$, the contribution $2J Y (1-Y)$ provides a boost (i.e. a gain) for the system's response, further stabilizing the states $Y=0$ (for low levels of input -i.e. $h<0$-, that is interpreted as {\em chemical noise} and it is thus suppressed) and $Y=1$ (for high level of input -i.e. $h>0$-, that is interpreted as a {\em chemical signal} and it is thus amplified). The usefulness of this Condorcet-like mechanism at work in biochemistry becomes evident already in the paradigmatic case of hemoglobin (a cooperative protein responsible for oxygen transport in tissues): when in the lungs (rich of oxygen), hemoglobin uses cooperativity to bind to as much oxygen as possible; when in the tissues (poor of oxygen) it uses cooperativity to get rid of it, thus releasing oxygen in the tissue.

\section*{Discussion}
In this manuscript we deepened our {\em translation} of  biochemical kinetics into a statistical mechanical scaffold started in \cite{ABBDU-SciRep2014,AABDK-SciRep2015} and that allows a straightforward cybernetic interpretation of these  phenomenologies. The final goal is to contribute in the quantitative understanding of the emergent computational properties possibly shown by large networks of biochemical reactions regarding cell signalling.
\newline
The route we paved is based on the one-to-one, robust and sharp, structural and behavioral analogy between the response functions in biochemical reactions (i.e. saturation curves), the response functions  of mean-field models in statistical mechanics (i.e. self-consistencies), and the response functions of key amplifiers in electronics (i.e. transfer functions). In particular, the response functions of cooperative (anticooperative) systems match those pertaining to mean-field ferromagnets (antiferromagnets), which, in turn, overlap those characterizing an operational amplifier (flip-flop).
\newline
We decided to keep the discussion at the mean-field level and in the canonical ensemble because it is within these limits that other theories regarding information processing systems (e.g., neural networks in Artificial Intelligence \cite{amit}) have been developed in the past, and it is by comparing our results to these theories that we aim to learn how information is treated during biochemical reactions. Notice that, from a statistical mechanical perspective, neural networks are particular types of spin-glasses, namely tricky realizations of ferromagnets and antiferromagnets broadly interacting, while from an electronic perspective, neural networks are realized by suitably combining large numbers of amplifiers and flip-flops, thus the first steps in this structural equivalence should focus on these basic ingredients, that have been indeed the subject of the present paper.
\newline
Moreover, many of the biochemical reactions of current interest involve small numbers of agents and this makes theories in the thermodynamic limit \cite{TDL} too coarse for the purpose of tackling the proposed analogy in full generality. Therefore, in this manuscript we tried to fix this point by developing a proper mathematical technique able to account  even for small-sized systems. \\
Summarizing the whole procedure, our route first requires mapping the original biochemical problem into a Hamiltonian formulation; the resulting model can then be embedded into a statistical mechanical framework. Next, the solution for this kind of systems is attainable by noting that their related free energies obey Hamilton--Jacobi type equations in the space of their parameters (ultimately mapping a problem in biochemistry into a problem of analytical mechanics): relying on the complete integrability of the system of multidimensional PDEs, we solve the general scenario in both the thermodynamic limit and the finite size case~\cite{Courant}. In particular, we observed that the multidimensional equations of state can be constructed via the hodograph equations that in turn provide the solution to a system of hydrodynamic type~\cite{Tsarev}.
Following this route we are able to provide a theoretical description of several complex phenomena stemming from finite-size effects. This phenomenology is rather broad and includes the effective cooperativity induced by stochasticity, as well as an enhanced bistability when dealing with toggles. Crucially, our procedure naturally allows a further interpretation of the response functions of these biochemical systems in terms of cybernetics, that constitutes a novel and transparent way to analyze how information is handled during these reactions: once shown the one-to-one behavioral correspondence between saturation functions in chemical kinetics, self-consistencies in statistical mechanics and transfer function in electronics (these are all identical saturable response functions from a cybernetic perspective), we related Condorcet phenomenology with signal amplification and noise suppression and framed it into the elementary scenario of {\em ferromagnetic gain}.
\newline
We tested our theory both against classical experiments (i.e., in the large $N$ limit), recovering all the main equations of reaction kinetics as suitable limits (i.e. Micaelis--Menten, Hill, Adair equations)  as well as against novel experiments involving small system sizes, recovering the expected phenomenology \cite{Lipshtat-PRL2006,Warren3,Gardner-Nature2000}.
\newline
Further developments of the theory now should proceed in three ways: on one side, still keeping the Maxwell-Boltzmann prescription, we can combine small (bio-chemical) circuits together in order to analyze information processing in larger chemical networks. On the other side, efforts are still needed to enlarge this scheme in order to apply to general out-of-equilibrium regimes (for instance with time-dependent field variations). Finally the above procedure can be further enriched by working in synergy with other well developed (and possibly alternative) mathematical methods, especially those already {\em thermodynamically oriented} -i.e., thougth to deal with the complexity of evaluating the partition function at finite size- as, for instance, those geometrically-oriented reported in the review by Ruppeiner \cite{ruppeiner}.

\section*{Acknowledgments}
Authors wish to thank the London Mathematical Society for the partial support to this research through the Scheme 4 Grant Ref. 41553 and INdAM (Istituto Nazionale di Fisica Mathematica) - GNFM (Gruppo Nazionale per la Fisica Matematica) for partial financial support through the Grant "Meccanica Statistica per il Deep Learning" and the Faculty of Engineering and Environment University of Northumbria Newcastle for supporting open access publication.
\newline
A.B. is grateful to LIFE group for partial financial support through {\em Programma INNOVA, Progetto MATCH}.
\newline
A.M. is grateful to the Department of Mathematics of University of Rome La Sapienza for the kind hospitality.
\newline
L.D.S. gratefully acknowledges funding through the CRC 1060 at the University of Bonn.


\vspace{1cm}

{\Large {\bf Author contribution statement} }\\

 All authors, Agliari E., Barra A., Dello Schiavo L., Moro A. have equally contributed to all phases of realisation of this manuscript in all its parts.

\vspace{1cm}

{\Large {\bf Additional Information}}\\

{\bf Competing Financial Interests statement} \\

The authors declare no competing financial interests.

\end{document}